\documentclass[12pt, hidelinks]{article}
\usepackage[utf8]{inputenc}
\usepackage{amsmath, amssymb, url, float, graphicx, float}
\usepackage{fullpage}
\usepackage[small,bf]{caption}
\usepackage{cite}

\newcommand{\tmin}{\theta^\mathrm{min}}
\newcommand{\tmax}{\theta^\mathrm{max}}

\newcommand{\ii}{\mathbf{i}}

\newcommand{\rp}{r_\mathrm{phys}}

\renewcommand\phi\varphi
\newcommand{\eps}{\varepsilon}

\newcommand{\ones}{\mathbf 1}
\newcommand{\reals}{{\mbox{\bf R}}}

\newcommand{\symm}{{\mbox{\bf S}}}  % symmetric matrices

\newcommand{\range}{{\mathcal R}}

\newcommand{\diag}{\mathop{\bf diag}}

\DeclareMathOperator*{\argmin}{argmin}

\newcommand{\sign}{\mathop{\bf sign}}

\newcommand{\cf}{{\it cf.}}
\newcommand{\eg}{{\it e.g.}}
\newcommand{\ie}{{\it i.e.}}
\newcommand{\etc}{{\it etc.}}

\newcommand{\BEAS}{\begin{eqnarray*}}
\newcommand{\EEAS}{\end{eqnarray*}}
\newcommand{\BEA}{\begin{eqnarray}}
\newcommand{\EEA}{\end{eqnarray}}
\newcommand{\BEQ}{\begin{equation}}
\newcommand{\EEQ}{\end{equation}}
\newcommand{\BIT}{\begin{itemize}}
\newcommand{\EIT}{\end{itemize}}

\title{Heuristic Methods and Performance Bounds for Photonic Design}
\author{Guillermo Angeris \\ {\small \texttt{angeris@stanford.edu}}
    \and Jelena Vu\v{c}kovi\'c \\ {\small \texttt{jela@stanford.edu}}
    \and Stephen Boyd\\ {\small \texttt{boyd@stanford.edu}}
}
\date{November 2020}

\begin{document} 
\maketitle 

\begin{abstract} In the photonic design problem, a
scientist or engineer chooses the physical parameters of a device to best
match some desired device behavior. Many instances of the photonic design problem
can be naturally stated as a mathematical optimization problem that is
computationally difficult to solve globally. Because of this, several heuristic
methods have been developed to approximately solve such problems.
These methods often produce very good designs, and, in many practical
applications, easily outperform `traditional' designs that rely on human intuition.
Yet, because these heuristic methods do not guarantee that the 
approximate solution found
is globally optimal, the question remains of just how much better a designer
might hope to do.  This question is addressed by performance bounds or 
impossibility results, which determine a performance level that
no design can achieve.  We focus on algorithmic performance bounds,
which involve substantial computation to determine.
We illustrate a variety of both heuristic methods and performance bounds
on two examples.  In these examples (and many others not reported here)
the performance bounds show that the heuristic designs are nearly optimal,
and can be considered globally optimal in practice. This review serves
to clearly set up the photonic design problem and unify existing approaches
for calculating performance bounds, while also providing some natural generalizations
and properties.

%
%, as a result, heuristics methods have been developed to approximately solve the problem. These heuristics yields designs that have good practical performance, but then there is the question of how much better could one do. This question is addressed by performance bounds, which state that, for a given design problem, [these are also called impossibility results because they state that certain specifications cannot be achieved. Traditional bounds are specified by gross features of a problem, and these give insight. More recently, people have investigated computational bounds, which are used to give a numerical certificate.
%    
%    These heuristic methods often produce very good design, and in many pracitcal applications, that is all that is needed. But the fact that these methods cannot guarantee that the solution found is global suggest the question of how much better one can do. This paper serves to clearly set up the problem, survey a variety of heuristic methods, and a variety of performance limits.
\end{abstract}

\section*{Introduction}
An important part of photonics, and many other scientific and engineering fields,
is the design and
construction of physical devices.  A `physical device', as used in this paper,
includes anything as simple as a spherical lens, where a scientist can easily
find an optimal lens for a given application by using basic algebra and ray
optics, all the way to potentially very complicated devices and applications
such as range detection and mapping using
LiDAR~\cite{yangInversedesignedNonreciprocalPulse2020}, where building an
`optimal' device, in nearly any practically useful sense, is an open research
problem. Though the two examples we give are in the field of photonics 
(and this is, indeed, the focus of our review), the
term refers to any device that can, at least theoretically, be built, and whose
desired behavior can be mathematically specified.

Traditionally, the design of physical devices was done by an engineer or
scientist, whom we will generally call a designer, for a specific application.
The designer would have a library, either physically or through experience, of
well-understood components or materials, each performing a specific function.
These components would then be carefully pieced together, often with a good
amount of ingenuity, in order to perform the desired task. In many cases, the
resulting designs could then be modified in part or in whole, or combined with
other designs, in order to perform even more complicated functions. This
procedure, while effective in practice, is time consuming
and sometimes even tangential to the final application of the design itself.

A second approach to constructing physical devices was initially explored in
the early 1960s within the field of electrical engineering, originally for the
purpose of recognizing printed
letters~\cite{kamentskyComputerAutomatedDesignMultifont1963} and has since been
extended to many other fields~\cite{
	jamesonAerodynamicDesignControl1988,
	jamesonOptimumAerodynamicDesign1998,
	leeDigitalFilterDesign1998,
	reutherConstrainedMultipointAerodynamic1999,
	nadarajahComparisonContinuousDiscrete2000,
    guanBridgeTopologyOptimisation2003,
	bendsoeTopologyOptimization2004,
	kaoMaximizingBandGaps2005,
	jiaoDemonstrationSystematicPhotonic2005,
    eharaTopologyDesignTensegrity2010,
    lalau-keralyAdjointShapeOptimization2013,
    ganapatiLightTrappingTextures2014,
    menRobustTopologyOptimization2014,
    callewaertInversedesignedStretchableMetalens2018,
    hughesAdjointMethodInverse2018,
    shiOptimizationMultilayerOptical2018,
    ahmedOptimizationThermalDesign2018,
    moleskyInverseDesignNanophotonics2018,
    wangAdjointbasedOptimizationActive2018,
    christiansenInverseDesignNanoparticles2020,
    michaelsHierarchicalDesignOptimization2020,
    linTopologyOptimizationFreeform2019}, with sometimes very surprising
results~\cite{thompsonEvolvedCircuitIntrinsic1997}.  In this approach, the
designer specifies a mathematical objective function which, given a design,
outputs a number representing how well the input design matches the desired
specifications; the lower this number is, the better the design.  This function
is then fed into an optimization algorithm, which attempts to minimize this
objective function by finding a design that is good in the sense specified by
the designer. In almost all practical cases, the algorithm will fail to find
the best possible design, and, except in very specific scenarios, may never do
so even when left to run for a very long time. But, in many applications, the
resulting designs found often have much better performance than any design
found by humans. This approach can be seen as a declarative approach to design:
the user specifies \emph{what} they want, while ceding control of \emph{how} it
should be done to the optimization algorithm. This idea has many names in
different fields, with field-specific connotations and denotations; these
include `automated design', `computational design', `inverse design' (in
photonics and aerospace engineering), `shape design' and `generative design'
(in mechanical engineering), `topology design' (in several fields), or
`synthesis' (in hardware design), among many others. We will simply call this optimization problem
the `physical design problem', with the understanding that many, if not all, of
the previously mentioned problems are instances of the physical design problem.

\section{The physical design problem}
The usual physical design problem can be formally stated in many ways. In this
review, we focus on a simple but general formulation, which,
as we show in this section, includes many important
problems in photonic design.

\subsection{Physics}
%Very broadly speaking, the inverse design problem begins with a physical theory that describes
%the relationship between the problem \emph{parameters}, which we will denote as $\theta$, and \emph{field variables},
%which we will denote as $z$. In photonic design, the physical theory connecting the field
%variables and the parameters is usually, but not always, described by Maxwell's equations. In this case, the field variables $z$ 
%represent, \eg, the electric and magnetic fields, while the parameters $\theta$ represent the material permittivities and material
%permeabilities of the design, chosen at each point in space.

The design problem starts with a physical theory that describes the
behavior of the \emph{field} (which we will write as $z$) under some
\emph{excitation} (which we will write as $b$).  The field $z$ and excitation
$b$ are vectors in some (typically infinite dimensional) vector space.  We
focus here on the case when the physical theory is linear, in which case we can
write the \emph{physics equation} as
\begin{equation}\label{eq:physical-equation}
    Az = b,
\end{equation}
where $A$ is a linear operator. Problems in physical design are
governed by physics equations such as Maxwell's equations, Helmholtz's
equation, the heat equation, the Schr\"odinger equation, among many others,
which are linear in many important applications, and therefore of the form
of~\eqref{eq:physical-equation}.

\paragraph{Electromagnetic wave equation.} For example, a common way of writing the
electromagnetic (EM) wave equation in terms of the electric field $E$ and the currents $J$, for a
monochromatic wave with angular frequency $\omega$
is~\cite[\S2]{joannopoulosPhotonicCrystalsMolding2008},
\[
-\nabla\times\nabla\times E + \omega^2\mu_0\eps E = -\ii \omega\mu_0 J,
\]
where $\eps$ denotes the permittivities at each point in space and $\mu_0$ is
the magnetic permeability, which we assume to be constant throughout space,
in this example. We can then make the following correspondences:
\begin{equation}\label{eq:maxwell}
\underbrace{(-\nabla\times\nabla\times \cdot + \omega^2\mu_0\eps)}_A \underbrace{E}_{z} = \underbrace{-\ii \omega\mu_0 J}_b,
\end{equation}
which naturally leads to an equation of the form of~\eqref{eq:physical-equation}.

\paragraph{Discretization.} 
We will work with an appropriate discretization of the field,
excitation, and physics equation~\eqref{eq:physical-equation}.
We will overload notation to use the same symbols for their
discretized versions.  In the sequel,
the field $z$ will be a vector in $\reals^n$, 
the excitation $b$ will be a vector
in $\reals^m$, and the linear operator $A$ will be a matrix in 
$\reals^{m \times n}$.  
The physics equation~\eqref{eq:physical-equation} is then a set of $m$
linear equations in $n$ scalar variables.  
Complex fields and excitations can
be reduced to the real case by separating them into their real and 
imaginary parts.

\paragraph{Solutions and simulations.} For a fixed $A$ and excitation $b$, we
will call any $z$ which satisfies~\eqref{eq:physical-equation} a
\emph{solution} of the physics equation. In general there can be a unique
solution, many solutions, or no solution. We will focus on the case when there
is a unique solution, \ie, $m=n$ and $A$ is invertible, so $z=A^{-1}b$.

We refer to computing the field $z=A^{-1}b$, given $A$ and $b$, as a
\emph{simulation}.  There are many simulation methods, including generic
methods for solving linear equations such as
sparse-direct methods~\cite{amestoyPerformanceScalabilityBlock2019,
chenAlgorithm887CHOLMOD2008} or iterative
methods~\cite{hestenesMethodsConjugateGradients1952,
saadGMRESGeneralizedMinimal1986}, and custom methods crafted specifically for the
particular physics equations~\cite{smithDomainDecompositionMethods1997,
urbanWaveletMethodsElliptic2008,
gillmanFastDirectSolvers2011}. We note that,
in practical photonic design, the resulting linear systems can be very large, with the number
of variables, $n$, often in the millions or tens of millions. Solving systems
in the upper end of this range often requires the use of large-scale
linear solvers~\cite{suNanophotonicInverseDesign2020}.

\paragraph{Approximate solutions and physics residual.} 
Some of the methods we will see work with approximate solutions of the physics
equation.  For any field vector $z$ and excitation $b$, we define the
\emph{physics residual} as $\rp = Az-b$.  A reasonable numerical measure of the
size of the residual is $\|Ax-b\|/\|b\|$, where $\| \cdot \|$ is a norm,
typically the Euclidean norm $\| \cdot \|_2$.  Simulations, especially those
that use iterative methods, produce fields with small physics residuals.

\paragraph{Modes.}
A simple trick can be used to represent the modes of a system as a solution
to~\eqref{eq:physical-equation}.  Suppose the original (discretized) physics
equation is $Hz=\lambda z$, where $\lambda$ is the eigenvalue and $z$ is an
associated mode.  Directly expressing this as $Az=b$ with $A=H-\lambda I$ and
$b=0$ yield a physics equation that is not invertible and has multiple
solutions (including, of course, $z=0$).  To fix a unique solution, we use a
linear normalization and insist that $c^Tz=1$, where $c\in \reals^n$ is some
nonzero vector.  We then represent the mode equation and normalization as
$Az=b$ with
\begin{equation}\label{eq:modes}
    A = \begin{bmatrix}
        H - \lambda I\\
        c^T
        \end{bmatrix}, \qquad  b = \begin{bmatrix}
        0_n\\
        1
    \end{bmatrix}.
\end{equation}
This has a unique solution, provided $c$ is not an eigenvector of $H$ with
eigenvalue $\lambda$ and that $\lambda$ is a simple eigenvalue. Note that the linear normalization
in~\eqref{eq:modes} differs from the usual choice of normalization, $\|z\|_2 =
1$, where $\|\cdot\|_2$ is the Euclidean norm.

\subsection{Design parameters}\label{sec:design-params}
In physical design, the
designer is able to change the system physics
equation~\eqref{eq:physical-equation}, by choosing some
parameters that affect the physics, \ie, $A$ and $b$.
Thus $A$ and $b$ depend on some \emph{design parameters} $\theta \in \reals^d$.
For example, in photonic design, $\theta$ is generally a variable that controls the permittivities inside
of the device.
We can then write the physics
equation~\eqref{eq:physical-equation} with explicit dependence on the design
parameters as
\begin{equation}\label{eq:parametrized-physics}
    A(\theta)z = b(\theta).
\end{equation}
When $A(\theta)$ is invertible we have $z(\theta) = A(\theta)^{-1}b(\theta)$; 
\ie, the field also depends (implicitly) on the design parameters $\theta$.

\paragraph{Affine physics design.}
In many practical cases (and in all of the examples we will show),
$A$ and $b$ are \emph{affine functions} of the design parameters; \ie,
\[
    A(\theta) = A_0 + \sum_{i=1}^d \theta_i A_i, \quad b(\theta) = b_0 + \sum_{i=1}^d \theta_i b_i,
\]
where $A_i \in \reals^{m\times n}$ and $b_i \in \reals^m$.
These $A_i$ are usually sparse matrices and vectors; that is, each design parameter
$\theta_i$ affects just a few entries of $A$ and $b$. Because equation~\eqref{eq:parametrized-physics}
is affine in $\theta$ when
holding $z$ fixed, and affine in $z$ when holding $\theta$ fixed, this type of
equation is sometimes called `bi-affine' or `multilinear' in $\theta$ and $z$.

\paragraph{Diagonal physics design.} A common and useful special case
of~\eqref{eq:parametrized-physics} is when $b(\theta)$ is a constant,
$b(\theta) = b_0$, and $A(\theta)$ can be written as
\begin{equation}\label{eq:diagonal-design}
    A(\theta) = A_0 + \diag(\theta),
\end{equation}
where $\diag(\theta)$ is a matrix whose diagonal entries contain
the elements of the vector $\theta$ and is zero elsewhere.
In other words, $A_i = E_{ii}$, where $E_{ii}$ is the matrix with only one nonzero 
entry (which is one), in the $i,i$ entry.
We will call this special case of~\eqref{eq:parametrized-physics} the
\emph{diagonal physics equation}.

\paragraph{EM wave equation in diagonal form.} A specific example of a diagonal
physics equation are Maxwell's equations~\eqref{eq:maxwell}, whenever a
designer is allowed to vary the permittivities. In this case, we can define
$\theta$ to be proportional to the permittivities (with proportionality
constant $\mu_0\omega^2$) such that the following correspondences can be made:
\[
(\underbrace{-\nabla\times\nabla\times \cdot}_{A_0} + \underbrace{\mu_0\omega^2\eps}_{\diag(\theta)}) \underbrace{E}_{z} = \underbrace{-\ii \omega\mu_0 J}_b.
\]
This correspondence results in an equation of the form of~\eqref{eq:diagonal-design}.
(We can also similarly write the more general case where the designer
is allowed to vary both the permittivities and permeabilities, in this form. See, \eg,
the appendix in~\cite{angerisComputationalBoundsPhotonic2019}.)

\paragraph{Low-rank updates.} Whenever $A(\theta)$ is sparse (which is very
common in practice), it is often possible to compute an explicit factorization
of $A(\theta)$ (\eg, the sparse Cholesky factorization, when $A(\theta)$ is
positive semidefinite) which makes evaluating $A(\theta)^{-1}b(\theta)$
inexpensive, after the factorization.
Additionally, whenever the matrices $A_i$ are also low rank (as in
the diagonal case~\eqref{eq:diagonal-design}, for example), updates to the factorization of $A(\theta)$
can be efficiently computed~\cite{chenAlgorithm887CHOLMOD2008}.
This implies that we can also efficiently
evaluate $A(\theta')^{-1}b(\theta')$, when only a small number of entries of
$\theta'$ differ from those of $\theta$, given the factorization for $A(\theta)$.

\paragraph{Parameter constraints.} In general, a designer has constraints on
the parameters that can be chosen. Because of this, we will define the
\emph{feasible parameter set}, $\Theta \subseteq \reals^d$, such that only
design parameters satisfying $\theta \in \Theta$ are feasible or valid. In many
applications, $\Theta$ is a hyperrectangle (or box) indicating that each
component of $\theta$ must lie in some interval given by $\tmin, \tmax \in
\reals^d$, \ie,
\[
    \Theta = \{\theta \in \reals^d \mid \tmin \le \theta \le \tmax\},
\]
where the inequalities are elementwise.

In photonic design, it is often not possible to vary the permittivities
along an interval but are instead allowed to be one of two possible values.
This leads to another common set of parameter constraints, where each component of
$\theta$ is constrained to be exactly one of two elements (we will call this
class of constraints \emph{Boolean constraints}):
\[
    \Theta = \{\theta \in \reals^d \mid \theta_i \in \{\tmin_i, \tmax_i\} ~ \text{for} ~ i=1, \dots, d \}.
\]
In this class of constraints, the total number of parameters that are feasible is $|\Theta| =
2^d$. 
The feasible parameter set $\Theta$ can also include
fabrication constraints such as minimum possible feature sizes, among
other possibilities~\cite{piggottFabricationconstrainedNanophotonicInverse2017},
but we will focus on the common cases of box or Boolean constraints.

\paragraph{Normalization.}\label{sec:normalization}
We can re-parametrize the design parameters to lie between $-1$ and $1$ (or any
other limits).  So, without loss of generality, we can always consider the upper
and lower bounds to be $\tmin = -\ones$ and $\tmax = \ones$, where $\ones$ is the
vector with all entries equal to one.  To do this, we introduce a new parameter $\delta \in \reals^d$ and
define
\[
\theta = \bar\theta + \rho\circ \delta,
\]
where $\bar\theta = (\tmax +\tmin)/2$ is the parameter midpoint,
while $\rho = (\tmax - \tmin)/2$ is the parameter radius, and $\circ$ denotes the
elementwise (Hadamard) product.
The constraint $\theta \in \Theta$ becomes $-\ones \leq \delta \leq \ones$,
where the inequalities are elementwise, in the box-constrained case,
or $\delta \in \{-1,1\}^d$ in the Boolean case.
We then have $A(\theta) = \tilde A(\delta)$, $b(\theta)=\tilde b(\delta)$,
with $\tilde A$ and $\tilde b$ affine, and
\[
\tilde A_0 = A(\bar\theta), \qquad \tilde b_0 = b(\bar\theta),
\]
while
\[
\tilde A_i = \rho_i A_i, \quad
\tilde b_i = \rho_i  b_i, \quad
i=1,\ldots, d.
\]

%\paragraph{Basic example.} A simple instance of~\eqref{eq:parametrized-physics} in photonics appears in Maxwell's equations~\eqref{eq:maxwell}. In this case, the designer is often allowed to change the permittivity, $\eps$, in some region by choosing a specific material in that region. For example, the designer could be allowed to choose either silicon or air as the material within some region, which, in turn, will change the permittivity (and  therefore the matrix $A$). In this case, then, the design parameters $\theta$ specify the permittivities, such that $\theta_i$ represents the permittivity of the material being used within region $i$.

%XXX: What basic examples (if any) should we include here?

\subsection{Optimization problem}\label{sec:optimization-problem}
The design objective is often written as a function
of the fields, specifying how well the resulting field matches the desired
objective. This objective function could specify the power in a given direction,
the field overlap (\ie, the inner product between the current field and a desired one),
or the total energy, all of which can be written as functions depending only on the field,
that the designer may wish to optimize. For example, the designer may wish to maximize the
power transmitted through a specific port of a device at a given
frequency~\cite{piggottInverseDesignDemonstration2015}, or to maximize the
focusing efficiency of a lens within a specific
region~\cite{chungHighNAAchromaticMetalenses2020,
bayatiInverseDesignedMetalenses2020}.  Finding a device that best matches this objective
can be directly phrased as a mathematical optimization problem.

\paragraph{Objective function.} In other words, we seek to find
a design whose field optimizes an
\emph{objective function} $f: \reals^n \to \reals
\cup\{+\infty\}$. The function's input is a field $z$ (generated
by some design $\theta \in \Theta$) and its output is a number that
specifies how good or bad this field is, or how well the field matches the
designer's specification. Without loss of generality, we will assume that a higher number is
worse (\ie, a designer wishes to minimize $f$), but we can just as well
maximize $f$ by, equivalently, minimizing its negative, $-f$. We
allow the objective function $f$ to take on infinite values to denote hard
constraints on the desired field: if $f(z) = + \infty$ for some field $z$, then
$z$ is not a feasible field.

%\paragraph{Physical constraints.} Of course, simply minimizing the function $f$ is potentially not very useful: it is
%also important to the problem that the fields being evaluated are physically
%realizable. 
%In other words, there should exist a physical design which actually supports these fields.
%As described in the previous section, this means that there exist some design parameters $\theta$ such that the field being evaluated, $z$, satisfies the physical equation~\eqref{eq:physical-equation}.

\paragraph{Problem statement.} We can then compactly write the problem that a
designer wishes to solve (or approximately solve), which we will call
the \emph{physical design problem}:
\begin{equation}\label{eq:main}
    \begin{aligned}
        & \text{minimize} && f(z)\\
        & \text{subject to} && A(\theta)z = b(\theta)\\
        &&& \theta \in \Theta,
    \end{aligned}
\end{equation}
where the problem variables are the fields $z \in \reals^n$ and design
parameters $\theta \in \reals^d$, while the problem data include the matrix
$A(\theta) \in \reals^{m \times n}$, the excitation $b(\theta) \in \reals^m$,
and the parameter constraint set $\Theta \subseteq \reals^d$. We will call the
special case of~\eqref{eq:main} where the physics equation is the diagonal
physics equation~\eqref{eq:diagonal-design} the \emph{diagonal
physical design problem}. (As a reminder, in photonic design, $\theta$ is usually
proportional to the permittivities, while $A(\theta)$ is the operator corresponding
to the electromagnetic wave equation~\eqref{eq:maxwell}.)

%XXX: As a comment on the problem, there is the implicit constraint that, for any $\theta$ you choose, there must exist a $z$ which solves the equation $A(\theta)z = b(\theta)$.

\paragraph{Problem attributes.} Problem~\eqref{eq:main} has several important
properties.
%First, we note that there is an (additional) implicit constraint
%when writing $A(\theta)z = b(\theta)$, which implies that any feasible design
%parameters $\theta \in \Theta$ must have a corresponding solution to the
%physical equation.
In many practical cases, the function $f$ is a
convex function and the set $\Theta$ is a convex set, which implies that
problem~\eqref{eq:main} is a convex problem in the variable $z$, when holding
$\theta$ fixed, and is a convex problem in $\theta$, when holding $z$ fixed.
This property leads to some useful heuristics for approximately solving
problem~\eqref{eq:main}; \cf,~\cite{luInverseDesignNanophotonic2010,
luNanophotonicComputationalDesign2013}. Additionally, problem~\eqref{eq:main} 
often has a smooth (differentiable) objective function $f$, while $\Theta$ can almost always be
represented as a number of smooth equality and inequality constraints, which
happens in many practical applications and all examples presented in this
review. In this case, we can apply general nonlinear optimization solvers such
as IPOPT~\cite{wachterImplementationInteriorpointFilter2006} directly to
problem~\eqref{eq:main}.

\paragraph{Computational hardness.} On the other hand, it is not difficult to
show that even finding a feasible design and field for problem~\eqref{eq:main}
is, in general, a computationally difficult problem
(\ie, it is NP-hard) even when the design parameters $\theta$ are unconstrained;
that is, even if $\Theta = \reals^n$. To do this, we will reduce the \emph{subset sum
problem}~\cite{karpReducibilityCombinatorialProblems1972}, a problem known to
be NP-hard, to an instance of~\eqref{eq:parametrized-physics}. This would imply that,
if we could efficiently solve problem~\eqref{eq:parametrized-physics}, then we could efficiently solve
the subset sum problem, which is widely believed to be computationally hard to solve.
(See, \eg,~\cite{aaronsonWhyPhilosophersShould2011} for a good overview of P vs.\ NP
and its implications.)

The subset sum
problem asks: given $c \in \reals^n$, is there a nonzero binary vector $x \in
\{0, 1\}^n$ such that $c^Tx = 0$? We will show that, given $c$, we can answer
this question by finding a field $z$ and a design $\theta$ that satisfy the
constraints of~\eqref{eq:main}, which will imply that, in general,
problem~\eqref{eq:main} is computationally difficult.

First, note that we can write the following conditions on $z$ and $\theta$,
\[
    c^Tz = 0, \quad \theta_{n+1}\ones^Tz = 1, \quad \theta_i(z_i - 1) = 0, \quad z_i = \theta_i, \quad i=1, \dots, n,
\]
as an instance of~\eqref{eq:parametrized-physics}, by appropriately stacking
the conditions into a matrix form. Now, the last two conditions are true if,
and only if, $z_i(z_i-1) = 0$, which also happens only when $z_i \in \{0, 1\}$
for $i=1, \dots, n$. The second condition implies that $\ones^Tz \ne 0$, and,
when combined with the first condition, this statement is true if, and only if,
there exists a nonzero solution to the subset sum problem, with the provided
vector $c$. This, in turn, shows that finding a feasible design $\theta$ and
field $z$ such that the constraints of~\eqref{eq:main} are satisfied must, in
general, be a problem that is at least as hard as the subset sum problem.

Because problem~\eqref{eq:main} is likely to be computationally difficult to
solve exactly when the number of parameters is large, we will focus on
heuristics which approximately solve the problem for the remainder of this
paper.

\paragraph{Multi-scenario design.}
A common design task is to find a single device that has good
performance across many different scenarios. For example, the device might need
to be robust against temperature variations, or the device might be required to
filter out a number of specific wavelengths, while allowing others through.
In this case, we will assume that the device
satisfies $S$ instances of the physics equation~\eqref{eq:physical-equation},
where the design parameters $\theta$ are held fixed across the $S$ instances,
but the physics equation or the excitation, is allowed to vary:
\[
    A^s(\theta)z^s = b^s(\theta), \quad s = 1, \dots, S.
\]
Here $A^s(\theta) \in \reals^{m_s \times n_s}$, $b^s(\theta) \in \reals^{m_s}$
and $z^s \in \reals^{n_s}$ for $s = 1, \dots, S$.  This leads to the
multi-scenario physical design problem:
\begin{equation}\label{eq:multi-scenario}
    \begin{aligned}
    & \text{minimize} && f(z^1, \dots, z^S)\\
    & \text{subject to} && A^s(\theta)z^s = b^s(\theta), \quad s=1, \dots, S\\
    &&& \theta \in \Theta,
    \end{aligned}
\end{equation}
where the variables are the fields $z^s \in \reals^{n_s}$ in each of the $s=1,
\dots, S$ scenarios and the design parameters $\theta \in \reals^d$, while the
objective function $f: \reals^{n_1}\times \dots \times \reals^{n_S} \to \reals$
can depend on the fields at any of the $S$ scenarios.

In fact, it turns out that we can write any instance of the multi-scenario
physical design problem~\eqref{eq:multi-scenario} as an instance
of~\eqref{eq:main}. To do this, we collect all of the individual physical
equations into a single constraint by placing all of the $A^s(\theta)$ along
the diagonal of a lager matrix $A(\theta)$, and stacking the excitations
$b^s(\theta)$ and fields $z^s$. More specifically, define
\[
    A(\theta) = \begin{bmatrix}
        A^1(\theta) & 0 & \dots & 0\\
        0 & A^2(\theta) & \dots & 0\\
        \vdots & \vdots & \ddots & \vdots\\
        0 & 0 & \dots & A^S(\theta)
        \end{bmatrix}, \quad b(\theta) = \begin{bmatrix}
        b^1(\theta)\\
        b^2(\theta)\\
        \vdots\\
        b^S(\theta)
        \end{bmatrix}, \quad z = \begin{bmatrix}
        z^1\\
        z^2\\
        \vdots\\
        z^S
    \end{bmatrix}.
\]
So we can write the $S$ distinct physical equations, $A^s(\theta)z^s =
b^s(\theta)$ for scenarios $s=1, \dots, S$, as the single equation, $A(\theta)z
= b(\theta)$, with dimensions $A(\theta) \in \reals^{m \times n}$, $z \in
\reals^n$, and $b(\theta) \in \reals^m$, where $m = m_1 + \dots + m_S$ and $n =
n_1 + \dots + n_S$. Overloading notation slightly, such that $f(z) = f(z_1,
\dots, z_S)$, then reduces problem~\eqref{eq:multi-scenario} to one of the form
of~\eqref{eq:main}. In other words, it suffices to only consider a problem of
the form of~\eqref{eq:main}.

\paragraph{Eliminating the field variables.} A simple (and relatively common)
equivalent formulation of problem~\eqref{eq:main} is to note that, because we
have assumed $A(\theta)$ is invertible, we can write it as a problem that
depends only on the design variables; \ie, since $z = A(\theta)^{-1}b(\theta)$,
we can write problem~\eqref{eq:main} as the following optimization problem over
$\theta \in \reals^d$:
\begin{equation}\label{eq:field-elimination}
    \begin{aligned}
    & \text{minimize} && f(A(\theta)^{-1}b(\theta))\\
    & \text{subject to} && \theta \in \Theta.
    \end{aligned}
\end{equation}
When the function $f$ is differentiable, we can easily compute its derivatives
with respect to each component of $\theta$:
\[
    \frac{\partial}{\partial\theta_i} f(A(\theta)^{-1}b(\theta)) = (\nabla f(z))^TA(\theta)^{-1}A_iz,
\]
where $z = A(\theta)^{-1}b(\theta)$ is the solution
to~\eqref{eq:parametrized-physics} for design parameters $\theta$. We can write
this in a slightly more compact form by defining $y = A(\theta)^{-T}(\nabla
f(z))$, such that
\begin{equation}\label{eq:adjoint}
    \frac{\partial}{\partial\theta_i} f(A(\theta)^{-1}b(\theta)) = y^TA_iz.
\end{equation}
Note that, to find $z$ and $y$, we only need to solve two systems of linear
equations, one over $A(\theta)$, and one over $A(\theta)^T$. This observation
can be used to efficiently compute the gradient of the objective with respect
to the design variables and is called the \emph{adjoint
method}, dating back to the control theory literature of the 1970s~\cite{lionsOptimalControlSystems1971,
gilesIntroductionAdjointApproach2000}. Other methods of computing
the derivative when, for example, the matrix $A(\theta)$ is not invertible
include automatic differentiation through the simulation
(as in~\cite{de2018end, hu2019difftaichi, minkovInverseDesignPhotonic2020}), among many
others.

\paragraph{Eliminating the design variables.} In the case of diagonal design,
and for some choices of the parameter constraint set $\Theta$, it is also
possible to eliminate the corresponding design
variables~\cite{angerisNewHeuristicPhysical2020}. For example, in the case
where the parameter constraint set $\Theta$ is a box, $-1 \le \theta_i \le
1$, then, given a field $z$, there exist design parameters $\theta$ satisfying
\[
    (A_0 + \diag(\theta))z = b_0,
\]
(\ie, $z$ satisfies the diagonal physics
equation~\eqref{eq:diagonal-design} with parameters $\theta$)
if, and only if the field $z$ satisfies
\[
    |A_0z - b_0| \le |z|,
\]
where the absolute value $|\cdot|$ is taken elementwise.

To see this, note that, from~\eqref{eq:diagonal-design}, we can write
\[
    A_0z - b_0 = -\diag(\theta) z,
\]
and, since we are free to choose any $-\ones \le \theta \le \ones$, we have
that such a $\theta$ exists, if, and only if,
\[
    |A_0z - b_0| \le |z|.
\]
This lets us write the diagonal physical design problem in the following
equivalent way:
\begin{equation}\label{eq:parameter-elimination}
    \begin{aligned}
    & \text{minimize} && f(z)\\
    & \text{subject to} && |A_0 z - b_0| \le |z|,
    \end{aligned}
\end{equation}
where the only problem variable is the field $z \in \reals^n$. We will call
this formulation of the diagonal physical design problem, as
in~\cite{angerisNewHeuristicPhysical2020}, the \emph{absolute-upper-bound}
formulation.

This rewriting also shows an interesting property of the physical design
problem, whenever $f$ is convex: if you know the signs of any optimal field,
then the design problem becomes convex and therefore easy to solve globally.
That is, if we know $s = \sign(z^\star)$, where $\sign$ is the signum function
\[
    \sign(z^\star)_i = \begin{cases}
        1 & z^\star_i \ge 0\\
        -1 & z^\star_i < 0,
    \end{cases}
\]
and where $z^\star$ is optimal for~\eqref{eq:parameter-elimination}, then any
solution to the convex optimization problem
\begin{equation}\label{eq:sign-parameter-elimination}
    \begin{aligned}
    & \text{minimize} && f(z)\\
    & \text{subject to} && |A_0z - b_0| \le \diag (s) z,
    \end{aligned}
\end{equation}
with variable $z \in \reals^n$ is a solution
to~\eqref{eq:parameter-elimination} with the same optimal value as $z^\star$.
This follows from two basic facts. First, any $z$ that is feasible
for~\eqref{eq:sign-parameter-elimination} is feasible
for~\eqref{eq:parameter-elimination} with the same objective value because
\[
    |A_0z - b_0| \le \diag(s) z \le |z|.
\]
And, second, that $z^\star$ is feasible
for~\eqref{eq:sign-parameter-elimination} since
\[
    \diag(s) z^\star = |z^\star|,
\]
and, by definition, $z^\star$ is feasible for~\eqref{eq:parameter-elimination}
and so satisfies $|A_0 z^\star - b_0| \le |z^\star|$.
This observation also leads to an optimization algorithm which
iteratively updates the signs and yields a sequence of feasible fields with
decreasing objective value~\cite[\S3]{angerisNewHeuristicPhysical2020}, called
sign-flip descent or SFD for short. We compare its performance against other
basic solvers in~\S\ref{sec:numerical-examples}.

A similar
analysis holds when $\Theta = \{-1, 1\}^n$, \ie, $\theta$ is constrained to be Boolean.
In this case, we have that, given some field $z$, there exists a Boolean design
$\theta \in \{-1,1\}^n$ such that
$\theta$ and $z$ satisfy the diagonal physics equation if, and only if,
$z$ satisfies
\begin{equation}\label{eq:boolean-parameter-elimination}
|A_0z - b_0| = |z|.
\end{equation}
We note that, in photonics, the box-constrained formulation sometimes leads to designs that cannot be practically
implemented because the resulting designs may have permittivities that lie along an interval (\ie, $\theta_i$ may lie
anywhere in $[-1, 1]$), while most
fabrication methods only allow the use of two possible permittivities (\ie, we must have $\theta_i \in \{-1, 1\}$). Despite this, the box-constrained formulation
is often used as a good initialization for current inverse design algorithms, which then approximately solve the Boolean case~\cite{suNanophotonicInverseDesign2020}.

%
%\paragraph{Computational hardness.} We note that problem~\eqref{eq:main} is, generally speaking, likely to be computationally hard to solve, even when the objective function
%$f$ is `nice' (\eg, smooth, convex, or even linear), because finding a feasible design and field for the constraint given by equation~\eqref{eq:parametrized-physics} is NP-hard, as is shown in~\S\ref{sec:design-params}. Because this is true, attempting to solve~\eqref{eq:main} exactly via global methods is likely to be computationally infeasible for anything but the smallest designs. We therefore focus on approximately solving~\eqref{eq:main} via heuristics that appear to be quite useful and give practical designs with very good performance in a relatively short amount of time.

\paragraph{Approximate solution methods.} There are many practical methods for
approximately solving~\eqref{eq:main}. For example, many of the earliest
solution methods approximately solve the physical design problem by applying
zeroth order (or `derivative free') optimization algorithms after eliminating
the field variable, as shown in
problem~\eqref{eq:field-elimination}~\cite{hauptComparisonGeneticGradientbased1995,
jervaseDesignResonantcavityenhancedPhotodetectors2000,
cormierRealcodedGeneticAlgorithm2001, kerrinckxPhotonicCrystalFiber2004,
sanchisIntegratedOpticalDevices2004, hakanssonInverseDesignPhotonic2005,
gohGeneticOptimizationPhotonic2007}.  Such methods are easy to implement in
practice, as they only require the ability to perform a basic simulation; \ie,
to solve the physics equation~\eqref{eq:parametrized-physics} for a given
$\theta \in \Theta$. Zeroth order optimization methods include hill-climbing,
genetic algorithms~\cite{hornbyAutomatedAntennaDesign2006}, simulated
annealing~\cite{kirkpatrickOptimizationSimulatedAnnealing1983},
Nelder-Mead~\cite{nelderSimplexMethodFunction1965}, and adaptive coordinate
descent~\cite{loshchilovAdaptiveCoordinateDescent2011}, among many
others~\cite{riosDerivativefreeOptimizationReview2013}.  While effective at
finding designs with moderate to good performance, zeroth order optimization
methods scale poorly and suffer from slow convergence when compared to
higher-order methods.
(See~\cite{connIntroductionDerivativeFreeOptimization2009}
and~\cite{audetDerivativeFreeBlackboxOptimization2017} for more information on
zeroth order optimization methods.)

A second important family of optimization algorithms, which include the
algorithms most used in practice, are the first order optimization algorithms,
which are also almost always applied to problem~\eqref{eq:field-elimination}.
In these cases, such methods additionally make use of gradient
information~\eqref{eq:adjoint}, leading to better computational
performance and faster convergence times, at the expense of higher
implementation complexity, as these methods require additional information from
the simulator. Examples of first order optimization algorithms used in practive
include L-BFGS-B~\cite{byrdLimitedMemoryAlgorithm1995}, proximal gradient
methods~\cite{parikhProximalAlgorithms2014}, and the method of moving
asymptotes~\cite{svanbergMethodMovingAsymptotes1987}, among many others.
(See~\cite{bertsekasNonlinearProgramming2016, nlopt,
kochenderferAlgorithmsOptimization2019} for a comprehensive overview.)

We compare a few different methods in~\S\ref{sec:small-numerical-examples}
in terms of computational performance and resulting design performance.

\section{Performance limits}
\label{sec:performance-limits}
Any approximate optimization method for the physical design
problem~\eqref{eq:main} can be used to generate approximately optimal designs.
In other words, if we let $p^\star$ be the optimal value for~\eqref{eq:main},
these procedures generate a design and field that satisfy the physics
equation~\eqref{eq:physical-equation} whose objective value, say $p$, satisfies
$p \ge p^\star$. In general, because problem~\eqref{eq:main} is hard to solve,
it is hard to know how far away our designs are from the true optimal value
$p^\star$. For example, once we have approximately optimized~\eqref{eq:main}
and received some design with objective value $p$, it is not clear if this
design is close to optimal (and no design can do significantly better) or if
there are designs that have much better performance than the one we've found.

It is an old tradition in physics to then ask: what is the best possible value
that we can hope to achieve? More specifically: is there some lower bound $d$
such that we can guarantee that the optimal value of problem~\eqref{eq:main},
$p^\star$, is never smaller than this bound; \ie, $p^\star \ge d$? Such a bound
can be interpreted in many ways.  For example, it can be interpreted as an
`impossibility result', which states that no device that satisfies the
physics equation~\eqref{eq:parametrized-physics} and the parameter constraints,
$\theta \in \Theta$ can have objective value smaller than $d$. We can also
interpret is as a `certificate of optimality': given some design with
objective value $p$, if $p$ is close to $d$, then $p$ must also be
close to the optimal objective value, $p^\star$, since $p \ge p^\star \ge d$.
Of course the best performance bound is $p^\star$, but computing this is 
intractable, and we seek bounds that can be computed at reasonable cost.

Additionally, performance bounds can be very important in speeding up the design process.
For example, it is often not clear how large a design needs to be in order
to achieve reasonable performance. This often results in designers having
to experiment with the total device size in order to find a design
which has at least the desired performance. Lower bounds on these values,
if they are efficiently computable, would give an indication of how large
a design needs to be in order to achieve the designer's goals
without additional (potentially very computationally expensive) experimentation.

\paragraph{Methods.} Roughly speaking, there are two main approaches to the
problem of finding lower bounds to the optimal objective value of
problem~\eqref{eq:main}. The first, and likely the earliest of the methods, is
to make basic physical assumptions about the system (for example, that the
system size is substantially smaller than the wavelength of the
excitation~\cite{wheelerFundamentalLimitationsSmall1947,
chuPhysicalLimitationsOmni1948}) and derive bounds on the corresponding
quantities~\cite{mckellarSumRulesOptical1982, yuFundamentalLimitNanophotonic2010,
yuThermodynamicUpperBound2012,millerPhotonExtractionKey2013,
miroshnichenkoUltimateAbsorptionLight2018}. While these methods are
historically important and yield good rule-of-thumb heuristics for design, many
of the bounds derived in this way require assumptions that are not
satisfied by the devices found by inverse design, or result in weak
bounds. The second approach, which has become relatively popular recently (see,
\eg,~\cite{millerFundamentalLimitsOptical2016, millerLimitsOpticalResponse2017,
    angerisComputationalBoundsPhotonic2019, moleskyBoundsAbsorptionThermal2019,
    shimFundamentalLimitsNearField2019,
    trivediBoundsScatteringAbsorptionless2020, moleskyGlobalOperatorBounds2020,
    moleskyHierarchicalMeanFieldOperator2020,
    moleskyTOperatorLimitsElectromagnetic2020,
moleskyFundamentalLimitsRadiative2020, michonLimitsSurfaceenhancedRaman2020,
kuangComputationalBoundsLightmatter2020, gustafssonUpperBoundsAbsorption2020,
chungHighNAAchromaticMetalenses2020}), essentially uses basic properties of the
constraints and objective function of problem~\eqref{eq:main} to derive bounds
on the best possible performance of the problem.  Such approaches include
algebraic manipulations of the physics equation~\eqref{eq:physical-equation}
combined with the parameter constraints $\theta \in \Theta$, and applications
of Lagrange duality to problem~\eqref{eq:main}. The resulting bounds often do
not have analytical forms, but can be numerically evaluated by an efficient
algorithm and are therefore called \emph{computational bounds}. We will discuss
such bounds in this section.

\subsection{Lagrange duality}
The basic tool in a number of these bounds is the use of Lagrange
duality. The idea is as follows. Given the optimization problem
\begin{equation}\label{eq:primal-problem}
    \begin{aligned}
        & \text{minimize} && f(x)\\
        & \text{subject to} && h(x) \le 0,
    \end{aligned}
\end{equation}
for some objective function $f: \reals^n \to \reals$, constraint function $h:
\reals^n \to \reals^m$, and optimization variable $x \in \reals^n$,
we form the \emph{Lagrangian}: 
\[
    L(x, \lambda) = f(x) + \lambda^Th(x),
\]
where $\lambda \in \reals^m$, with $\lambda \ge 0$ is a Lagrange multiplier
or dual variable. This lets us define the \emph{Lagrange
dual function} $g: \reals^m \to \reals$, given by
\[
    g(\lambda) = \inf_x L(x, \lambda) = \inf_x \left(f(x) + \lambda^Th(x)\right).
\]
See, \eg,~\cite[\S5]{cvxbook} for more information on Lagrange dual functions.

\paragraph{Lower bound property.} The function $g$ has a few interesting
properties. First, for any $\lambda \ge 0$, $g(\lambda)$ is always smaller than
$p^\star$, the optimal value of~\eqref{eq:primal-problem}, and is therefore a performance
bound.  More specifically, we have
\[
    g(\lambda) = \inf_x \left(f(x) + \lambda^Th(x)\right) \le \inf_{h(x) \le 0} \left(f(x) + \lambda^Th(x)\right) \le \inf_{h(x) \le 0} f(x) = p^\star.
\]
The first inequality follows from the fact that the set of $x$ which satisfy
$h(x) \le 0$ is no larger than the set of all $x \in \reals^n$, and the second
follows from the fact that, because $h(x) \le 0$ and $\lambda \ge 0$, then
$\lambda^Th(x) \le 0$.
Thus the Lagrange dual function gives us a performance bound, parametrized
by $\lambda$.
(Depending on the problem and choice of $\lambda$, it can give the 
trivial lower bound $-\infty$.)

\paragraph{Concavity.} Since $g(\lambda)$ is a performance bound for any
$\lambda \ge 0$, it is then natural to ask, what is the best possible performance
bound? In other words, what is the largest possible value of $g(\lambda)$ over
the possible values of $\lambda$? This problem is called the
\emph{dual problem} and can be written as
\begin{equation}\label{eq:dual-problem}
    \begin{aligned}
        & \text{maximize} && g(\lambda)\\
        & \text{subject to} && \lambda \ge 0.
    \end{aligned}
\end{equation}
In general, evaluating $g(\lambda)$ at some $\lambda \ge 0$ is at least
as hard as solving the original problem~\eqref{eq:primal-problem}.  On the
other hand, when it is possible to efficiently evaluate $g(\lambda)$, it is
almost always possible to efficiently find the optimal value of the dual
problem~\eqref{eq:dual-problem} because the function $g$ is always a concave
function, even when the objective function $f$ and constraints $h$ in the
original problem are not convex~\cite[\S5.1.2]{cvxbook}.

\paragraph{Initializations.} A solution to the dual problem~\eqref{eq:dual-problem}
often suggests a good initialization for heuristics which attempt to minimize
problem~\eqref{eq:primal-problem}. Given some dual variable $\lambda^\star$
that is optimal for~\eqref{eq:dual-problem}, there exists some $x^0$ which minimizes
the Lagrangian at this choice of dual variable
\[
x^0 \in \argmin_x \,L(x, \lambda^\star),
\]
under some basic assumptions on the objective function $f$ and constraints $h$.
In practice, $x^0$ is generally close to a reasonable design (see, \eg,~\cite{angerisComputationalBoundsPhotonic2019})
even if it is not feasible; \ie, it need not satisfy $h(x^0) \le 0$, except in some special
scenarios such as when the functions $f$ and $h$ are both convex, in which case $x^0$ is globally
optimal~\cite[\S5.2]{cvxbook}. Because it is often true that $x^0$ is easy to evaluate
whenever $g(\lambda^\star)$ is easy to evaluate, this initialization can be seen
as a by-product of finding a solution to~\eqref{eq:dual-problem}.

\subsection{Local power conservation}
\label{sec:local-power-conservation}
The first approach to constructing bounds for~\eqref{eq:main} was presented
originally in some generality in~\cite{millerFundamentalLimitsOptical2016} and
then~\cite{moleskyFundamentalLimitsRadiative2020} and later extended and fully
clarified in~\cite{gustafssonUpperBoundsAbsorption2020,
kuangMaximalSinglefrequencyElectromagnetic2020,
moleskyGlobalOperatorBounds2020} and subsequently fully generalized
in~\cite{moleskyHierarchicalMeanFieldOperator2020}
and~\cite{kuangComputationalBoundsLightmatter2020} in the case where the parameters
are Boolean (\ie, $\Theta = \{-1, 1\}^n$). We will present a further generalization
to the case where the parameters are box-constrained ($\Theta = [-1, 1]^n$), by a slightly
different proof that considers a relaxation of the absolute-upper-bound
formulation~\eqref{eq:parameter-elimination}. (The
Boolean case follows from an identical argument by
considering~\eqref{eq:boolean-parameter-elimination},
instead.)
Bounds of this form are
essentially power conservation laws over a given subdomain, which,
in photonics, are often included under the name `optical
theorem'~\cite{newtonOpticalTheorem1976}.

\paragraph{Power inequalities.} Starting with the absolute-upper-bound formulation given in~\eqref{eq:parameter-elimination},
we can square both sides of the inequality constraint to receive an equivalent formulation,
\begin{equation}\label{eq:aub-quadratic-problem}
\begin{aligned}
& \text{minimize} && f(z)\\
& \text{subject to} && (a_i^T z - (b_0)_i)^2 \le z_i^2, \quad i=1, \dots, n.
\end{aligned}
\end{equation}
In other words, a field $z$ is feasible (\ie, there exists a design $\theta \in \Theta$ such that
$z$ and $\theta$ satisfy the diagonal physics equation~\eqref{eq:diagonal-design})
if, and only if the following quadratic inequalities are all satisfied:
\begin{equation}\label{eq:aub-quadratic}
(a_i^T z - (b_0)_i)^2 \le z_i^2, \quad i=1, \dots, n.
\end{equation}
We can, of course, multiply these inequalities by a nonnegative value $\lambda_i \ge 0$
for $i=1, \dots, n$
and add any number of them together to get another valid, quadratic inequality,
\[
\sum_{i=1}^n \lambda_i(a_i^T z - (b_0)_i)^2 \le \sum_{i=1}^n \lambda_i z_i^2,
\]
where $a_i^T$ is the $i$th row of $A_0$. This can be more compactly expressed as
\begin{equation}\label{eq:quadratic-inequality}
(A_0z - b_0)^TD(A_0z - b_0) \le z^TDz,
\end{equation}
where $D = \diag(\lambda)$ and has nonnegative entries along the diagonal.

The family of inequalities~\eqref{eq:quadratic-inequality}, parametrized by the
nonnegative diagonal matrices $D$ is the same as the family given
in~\cite{moleskyHierarchicalMeanFieldOperator2020,
kuangComputationalBoundsLightmatter2020}, except in the case where $\Theta$
specifies a box constraint, instead of a Boolean one.  Additionally, because
$z$ satisfies~\eqref{eq:aub-quadratic} if, and only if, it
satisfies~\eqref{eq:quadratic-inequality} for all diagonal matrices $D$ with
nonnegative diagonals, then it follows that the family of quadratic
inequalities in~\cite{moleskyHierarchicalMeanFieldOperator2020,
kuangComputationalBoundsLightmatter2020} is tight, in the following sense: any
field $z$ which satisfies these power conservation laws for all nonnegative
diagonal matrices $D$ must also have a corresponding design $\theta \in \Theta$
such that $z$ and $\theta$ simultaneously satisfy the diagonal physics
equation~\eqref{eq:diagonal-design}. (The inequalities
in~\eqref{eq:quadratic-inequality} are of a slightly different form than those
presented in~\cite{moleskyHierarchicalMeanFieldOperator2020,
kuangComputationalBoundsLightmatter2020}.  We show their equivalence in
appendix~\ref{app:bound-equivalence}.)

\paragraph{Relaxed formulation.} Because any feasible field $z$
satisfies~\eqref{eq:quadratic-inequality} for any nonnegative diagonal matrix
$D$, we can relax the family of inequalities~\eqref{eq:quadratic-inequality}
from all nonnegative diagonal matrices $D$ to a finite number of them, which we
will write as $D_j$ for $j=1, \dots, N$.  The following problem can then be
seen as a relaxation of~\eqref{eq:aub-quadratic-problem}:
\begin{equation}\label{eq:quad-relaxation}
\begin{aligned}
& \text{minimize} && f(z)\\
& \text{subject to} && (A_0z - b_0)^TD_j(A_0z - b_0) \le z^TD_jz, \quad j=1, \dots, N,
\end{aligned}
\end{equation}
with the field $z \in \reals^n$ as the only variable and problem data $A_0 \in
\reals^{n\times n}$, the excitation $b_0 \in \reals^n$, and the diagonal
matrices $D_j \in \reals^{n\times n}$ with nonnegative entries along the
diagonal, for $j=1, \dots, N$. Note that, if $D_j = E_{jj}$ for $j=1, \dots,
n$, where $E_{jj}$ is the matrix with a single nonzero in the $j,j$ entry
(which is one) and is zero elsewhere, we recover the original
problem~\eqref{eq:aub-quadratic-problem}. Additionally, while this problem is a
relaxation of the original, it is still likely to be computationally hard to
solve.

\paragraph{Quadratic objective.} 
In general, finding lower bounds to the relaxed
formulation~\eqref{eq:quad-relaxation} also need not be easy, except in some
important special cases. For example, when the objective function $f$ is a
quadratic,
\[
    f(z) = \frac12 z^TPz + q^Tz + r,
\]
for some symmetric matrix $P \in \symm^n$, vector $q \in \reals^n$, and $r \in
\reals$, then problem~\eqref{eq:quad-relaxation} is a quadratically constrained
quadratic program (QCQP, see~\cite{parkGeneralHeuristicsNonconvex2017}), and
the dual function $g$ corresponding to this problem has a closed form solution.
In fact, the dual problem of~\eqref{eq:quad-relaxation} is, in general, a
convex semidefinite program (SDP) which can be solved in practice for moderate
values of $n$ and $N$~\cite[\S1]{cvxbook}, allowing us to find a lower bound
to~\eqref{eq:quad-relaxation} and therefore to~\eqref{eq:main} efficiently.
 
\paragraph{Dual problem.} To find the dual problem, we first formulate the
Lagrangian of~\eqref{eq:quad-relaxation}:
\[
    \begin{aligned}
        L(z, \lambda) &= \frac12 z^T P z +  q^Tz + r + \frac12\sum_{j=1}^N \lambda_j((A_0z - b_0)^TD_j(A_0z - b_0) - z^TD_jz)\\
                      &= \frac12 z^TT(\lambda)z + v(\lambda)^Tz + u(\lambda),
    \end{aligned}
\]
where $\lambda \in \reals^N_+$ and we have defined
\[
    T(\lambda) = P + \sum_{j=1}^N \lambda_j (A^T_0D_jA_0 - D_j), ~~ v(\lambda) = q - \sum_{j=1}^N \lambda_j (A^T_0D_jb_0) , ~~ u(\lambda) = \frac12 \sum_{j=1}^N \lambda_j b^T_0D_jb_0 +r,
\]
such that $T(\lambda) \in \symm^n$ and $v(\lambda) \in \reals^n$, for
notational convenience. (This follows from expanding the expression and
collecting the quadratic, linear, and constant terms in the variable $z$.)

It is not hard to show that the dual function is, for $\lambda \ge 0$,
\[
    g(\lambda) = \inf_z L(y, \lambda) = \begin{cases}
        u(\lambda) -\frac12 v(\lambda)^TT(\lambda)^+v(\lambda) & T(\lambda) \ge 0, ~~ v(\lambda) \in \range(T(\lambda))\\
        - \infty & \text{otherwise}.
    \end{cases}
\]
(See, \eg,~\cite[\S3.2]{parkGeneralHeuristicsNonconvex2017}.) Here
$T(\lambda)^+$ is the Moore-Penrose pseudoinverse~\cite[\S11.5]{linalg} of
$T(\lambda)$, while $T(\lambda) \ge 0$ means that $T(\lambda)$ is positive
semidefinite, and $\range(T(\lambda))$ is the range of $T(\lambda)$. The
corresponding problem of maximizing $g$ over $\lambda \in \reals^N_+$ can be
written as a standard form SDP, which we will call the \emph{power dual bound}:
\begin{equation}\label{eq:dual-problem-family-bounds}
    \begin{aligned}
        & \text{maximize} && u(\lambda) - (1/2)t\\
        & \text{subject to} && \begin{bmatrix}
            t & v(\lambda)^T\\
            v(\lambda) & T(\lambda)
        \end{bmatrix} \ge 0\\
        &&& \lambda \ge 0,
    \end{aligned}
\end{equation}
with variables $t \in \reals$ and $\lambda \in \reals^N$. This problem can be
easily specified using domain-specific languages such as CVXPY or JuMP.jl and
solved using convex solvers that support SDPs, such as Mosek~\cite{mosek},
SCS~\cite{odonoghueConicOptimizationOperator2016}, or
COSMO.jl~\cite{garstkaCOSMOConicOperator2019}.  Because of the sparsity of
$A_0$, it is often the case that such problems have chordal
structure~\cite{vandenbergheChordalGraphsSemidefinite2015}, which can be
exploited to efficiently solve problem~\eqref{eq:dual-problem-family-bounds}
by some solvers such as COSMO.jl.

\paragraph{Optimal choice of $D_j$ matrices.} Given the dual
problem~\eqref{eq:dual-problem-family-bounds} of
problem~\eqref{eq:quad-relaxation}, the question remains of how to best choose
the diagonal matrices $D_j$ for $j=1, \dots, N$ such that the optimal value
of~\eqref{eq:dual-problem-family-bounds} is maximized. Historically, the
initial bounds in~\cite{millerFundamentalLimitsOptical2016,
    moleskyFundamentalLimitsRadiative2020, gustafssonUpperBoundsAbsorption2020,
kuangMaximalSinglefrequencyElectromagnetic2020,
moleskyGlobalOperatorBounds2020} assumed a number of fixed diagonal matrices
$D_j$.  Later, \cite{moleskyHierarchicalMeanFieldOperator2020}
and~\cite{kuangComputationalBoundsLightmatter2020} generalized the approach to
include any diagonal matrix $D$
and~\cite{kuangComputationalBoundsLightmatter2020} proposed an iterative
algorithm which, starting with some diagonal matrix $D_1$ (such as $D_1 = I$)
would solve a problem similar to~\eqref{eq:dual-problem-family-bounds} at
iteration $k$ and propose a new diagonal matrix $D_{k+1}$ that would be
appended to the constraints of~\eqref{eq:quad-relaxation}. This method is
conceptually similar to the cutting-plane method described
in~\cite[\S3.5]{parkGeneralHeuristicsNonconvex2017} applied to the dual
problem~\eqref{eq:dual-problem-family-bounds} with $D_j = E_{jj}$ for $j=1,
\dots, n$.  (More accurately, the procedure proposed
in~\cite{kuangComputationalBoundsLightmatter2020} solves an SDP relaxation
of~\eqref{eq:quad-relaxation} instead of the dual
problem~\eqref{eq:dual-problem-family-bounds}, but it can be shown that both
problems have the same optimal value by strong
duality~\cite[\S3.3]{parkGeneralHeuristicsNonconvex2017}.)

It is also reasonable to ask: what are the best possible choices of $D_j$ such
that the optimal value of~\eqref{eq:dual-problem-family-bounds} is maximized?
It is not hard to show that a single (correctly chosen) diagonal matrix, $D =
\diag(\lambda)$, suffices and that this matrix can be efficiently found. To see
this, note that we can choose $D_j =E_{jj}$ for $j=1, \dots, n$ such that,
\begin{equation}\label{eq:definition-d}
    D = \sum_{j=1}^n \lambda_j D_j = \diag(\lambda).
\end{equation}
This would let us write
\[
    T(\lambda) = P + A^T\diag(\lambda)A - \diag(\lambda), \quad v(\lambda) = q - A^T\diag(\lambda)b , \quad u(\lambda) = \frac12 b^T\diag(\lambda)b +r.
\]
We then note that picking $\lambda^\star$ optimal
for~\eqref{eq:dual-problem-family-bounds}, when it exists, gives a diagonal
matrix $D^\star = \diag(\lambda^\star)$. Additionally,
solving~\eqref{eq:quad-relaxation} with this choice of matrix $D^\star$ is a
special case where the number of quadratic constraints, $N$, equals $1$, which
can be efficiently solved and has the same optimal value
as~\eqref{eq:dual-problem-family-bounds}. (See, \eg, appendix B
of~\cite{cvxbook}.)

%such that
%solving~\eqref{eq:diag-lower-bound} using this single constraint (where the
%equality is relaxed to an inequality, $\le$) gives the same optimal value by
%strong duality. In fact, this bound is the best possible bound of this form,
%since any linear combination of diagonal matrices is, of course, a diagonal
%matrix. This also lets us see the algorithm of~\cite{kuangComputationalBoundsLightmatter2020} as a method for dual
%ascent, where, at iteration $k$, the matrix $(\diag\lambda)$ in~\eqref{eq:definition-d} is
%constrained to be written as a linear combination of the diagonal matrices
%$D_1, \dots, D_k$ proposed in the previous $k$ iterations. Based on the result
%at the $k$th iteration, a new matrix $D_{k+1}$ is then proposed and added to
%the list until a stopping condition is met.

%\paragraph{Extensions.} It is possible to extend the power conservation bounds
%to also include the case where the constraint set is a box,
%$\Theta = [-1, 1]^n$ rather than a Boolean constraint set.
%In this case, $z$ satisfies
%\begin{equation}\label{eq:box-power-conservation}
%z^T(A^TDA - D)z \le 2z^TA^TDb - b^TDb,
%\end{equation}
%for any diagonal matrix $D \in \reals^{n\times n}$ with nonnegative entries,
%if, and only if, there exists a $\theta \in [-1, 1]^n$ such that $z$ and $\theta$
%both satisfy the diagonal physics equation~\eqref{eq:diagonal-design}.
%While both the forward and backward implication use similar ideas as the
%bound in~\eqref{eq:family-bounds}, we give a proof of~\eqref{eq:box-power-conservation}
%in appendix~\ref{XXX}.

\subsection{Diagonal physics dual}
Another approach to computing lower bounds for the diagonal physical design
problem~\eqref{eq:main} is by a direct application of Lagrange duality,
originally given in~\cite{angerisComputationalBoundsPhotonic2019} and later
extended in~\cite{trivediBoundsScatteringAbsorptionless2020}
and~\cite{zhaoMinimumDielectricResonatorMode2020}. This approach gives a lower
bound when the objective function is separable:
\[
    f(z) = \sum_{i=1}^n f_i(z_i),
\]
and the constraints are of the following form:
\begin{equation}\label{eq:constraint-set}
    \Theta = \{\theta \mid -\ones \le \theta \le \ones, ~ \theta_i = \theta_j ~ \text{for} ~ i, j \in S_k, ~ k=1, \dots K\},
\end{equation}
where $S_k \subseteq \{1, \dots, d\}$ for $k=1, \dots, K$ are disjoint sets,
specifying which entries of $\theta$ are constrained to be equal.
% Not sure if we should write this
% We will assume that $\tmin_i = \tmin_j$ and $\tmax_i = \tmax_j$ for $i, j \in
% S_k$ for all $k=1, \dots, K$ (since we can always assume that $\tmin_i =
% \max_{j \in S_k} \tmin_j$ when $i \in S_k$.
We additionally note that a lower bound for this constraint set also yields a
lower bound for the Boolean case, since this constraint set contains the Boolean one.

\paragraph{Problem Lagrangian.} The basic trick here is to rewrite
problem~\eqref{eq:main} as the following (equivalent) problem:
\begin{equation}\label{eq:main-indicator}
    \begin{aligned}
        & \text{minimize} && {\textstyle \sum_{i=1}^n f_i(z_i) + I(\theta)}\\
        & \text{subject to} && (A_0 + \diag(\theta))z = b_0,
    \end{aligned}
\end{equation}
where we have pulled out the constraint $\theta \in \Theta$ into the indicator
function $I: \reals^n \to \reals \cup \{+\infty\}$ for the set $\Theta$,
\[
    I(\theta) = \begin{cases}
        0 & \theta \in \Theta\\
        +\infty & \theta \not\in\Theta.
    \end{cases}
\]
We can then easily formulate the Lagrangian of this problem:
\[
    L(z, \theta, \nu) = \sum_{i=1}^n f_i(z_i) + I(\theta) + \nu^T(A_0 + \diag(\theta))z - \nu^Tb_0,
\]
which we note is separable in terms of $z$:
\[
    L(z, \theta, \nu) = \sum_{i=1}^n \left(f_i(z_i) + (a_i^T\nu)z_i + \nu_i\theta_iz_i\right) + I(\theta) - \nu^Tb_0
\]

\paragraph{Dual function.} As before, the dual function is defined as
\[
    g(\nu) = \inf_{\theta}\inf_{z} L(z, \theta, \nu).
\]
Computing the inner infimum is relatively simple, which gives:
\[
    \begin{aligned}
        \inf_z L(z, \theta, \nu) &= \sum_{i=1}^n \,\inf_{z_i}\left( f_i(z_i) + (a_i^T\nu)z_i + \nu_i\theta_iz_i\right) + I(\theta) - \nu^Tb_0\\
                                     &= -\sum_{i=1}^n f_i^*(-a_i^T\nu - \nu_i\theta_i) + I(\theta) - \nu^Tb_0.
    \end{aligned}
\]
Here, $f_i^*: \reals \to \reals \cup \{+\infty\}$ is the \emph{convex
conjugate} of $f_i$ (sometimes called the Fenchel conjugate function or simply
the conjugate function) defined as
\[
    f_i^*(u) = \sup_x \left(ux - f_i(x)\right),
\]
and is well known for a number of functions~\cite[\S3.3]{cvxbook}.
Additionally, we will make use of the fact that $f_i^*$ is always a convex
function, even when $f_i$ is not convex.

To compute the outer infimum, we note that we can write
\[
    \begin{aligned}
        g(\nu) &= \inf_\theta \left(-\sum_{i=1}^n f_i^*(-a_i^T\nu - \nu_i\theta_i) + I(\theta) - \nu^Tb_0\right)\\
               &= \inf_\theta \left(-\sum_{k=1}^K \sum_{i \in S_k} f_i^*(-a_i^T\nu - \nu_i\theta_k) + I(\theta)\right) - \nu^Tb_0 \\
               &= -\sum_{k=1}^K \left(\sup_{-1 \le \theta_k \le 1} \sum_{i \in S_k} f_i^*(-a_i^T\nu - \nu_i\theta_k)\right) - \nu^Tb_0 \\
               &= -\sum_{k=1}^K \max \left\{\sum_{i \in S_k} f_i^*(-a_i^T\nu + \nu_i), ~\sum_{i \in S_k} f_i^*(-a_i^T\nu - \nu_i)\right\} - \nu^Tb_0,
    \end{aligned}
\]
where we have used the fact that $-\inf_x h(x) = \sup_x -h(x)$ for any function
$h$, and the fact that a scalar convex function achieves its
maximum over an interval at the boundary of that interval.

\paragraph{Dual problem.} Given the dual function $g$, we can find a lower
bound to the original problem by evaluating $g$ for any $\nu \in \reals^n$. We
can then ask what's the best possible dual bound, which gives the following
dual problem, which we will call the \emph{diagonal dual bound}:
\begin{equation}\label{eq:diagonal-dual}
    \begin{aligned}
    & \text{maximize} && -\sum_{k=1}^K \max \left\{\sum_{i \in S_k} f_i^*(-a_i^T\nu + \nu_i), ~\sum_{i \in S_k} f_i^*(-a_i^T\nu - \nu_i)\right\} - \nu^Tb_0.
    \end{aligned}
\end{equation}
This problem is a convex optimization problem with variable $\nu \in \reals^n$,
whose optimal value, $d^\star$, can almost always be efficiently found whenever
the function $f^*$ can be efficiently evaluated.

\paragraph{Field bounds.} In some cases, the bounds given
by~\eqref{eq:diagonal-dual} can sometimes be very weak; \eg,
if $f_i = 0$ for all indices $i$ except one. One way of improving
the lower bounds is to add a redundant constraint to~\eqref{eq:main-indicator};
\ie, a constraint such that, if $z$ satisfies $(A_0 + \diag(\theta))z = b_0$,
then it also satisfies $h(z) \le 0$ for some function $h: \reals^n \to \reals$,
such that the dual function of the resulting problem is still simple to
evaluate.

One useful example, originally presented
in~\cite{trivediBoundsScatteringAbsorptionless2020}, is to note that, if $z$
satisfies
\[
    (A_0 + \diag(\theta))z = b_0,
\]
then it also satisfies,
\[
    \|A_0z\|_2 - \|\diag(\theta) z\|_2 \le \|(A_0 + \diag(\theta))z\|_2 = \|b_0\|_2,
\]
where we have taken the norm of both sides of the expression, and the inequality
follows from the triangle inequality. Using the fact that 
\[
    \|A_0z\|_2 \ge \sigma_1(A_0)\|z\|_2 \quad \text{and} \quad \|\diag(\theta) z\|_2 \le \|z\|_2,
\]
where $\sigma_1(A_0)$ is the smallest singular value of $A_0$, we find
\[
    \sigma_1(A_0)\|z\|_2 - \|z\|_2 \le \|b_0\|_2,
\]
or, after some rearrangement:
\[
    \|z\|_2 \le \frac{\|b_0\|_2}{\sigma_1(A_0) - 1},
\]
whenever $\sigma_1(A_0) > 1$. Evaluating the dual for this new
problem is a simple extension of the procedure given above as
in~\cite[\S6]{angerisComputationalBoundsPhotonic2019}. In fact, the
same procedure can be extended to any norm $\|\cdot\|$ by replacing
$\sigma_1(A_0)$ with $1/\|A_0^{-1}\|$, where $\|A_0^{-1}\|$ is the
induced operator norm of $A_0^{-1}$:
\[
\|A_0^{-1}\| = \sup_{\|x\| = 1} \|A_0^{-1}x\|,
\]
but the resulting dual function need not be easy to evaluate.

\paragraph{Mode volume.} Another extension, presented originally
in~\cite{zhaoMinimumDielectricResonatorMode2020} is for an objective function
$f$ of the form
\[
f(z) = \frac{\|z\|^2_2}{z_i^2},
\]
whenever $z_i \ne 0$ and is $+\infty$ otherwise. Here, $i$ is a fixed index and
we will assume there is no excitation; \ie, $b_0 = 0$.  Note that this
objective function is similar to, and can be easily extended to include, the
cavity mode volume. Because $b_0=0$, then the physics equation and objective
are 0-homogeneous in $z$, so any feasible point $z$ with $z_i \ne 0$ can be
scaled by a nonzero value $\eta \in \reals$, such that $\eta z$ is also
feasible with the same objective value, $f(\eta z) = f(z)$. We can then fix a
normalization by setting $z_i = 1$ to get an equivalent problem:
\begin{equation}\label{eq:mode-volume}
    \begin{aligned}
        & \text{minimize} && \|z\|_2^2\\
        & \text{subject to} && (A_0 + \diag(\theta))z = 0\\
        &&& z_i = 1\\
        &&& -\ones \le \theta \le \ones,
    \end{aligned}
\end{equation}
with variables $z$ and $\theta$ and some fixed index $i$. The dual for problem~\eqref{eq:mode-volume} can be computed using the same
method presented in this section.

\section{Numerical examples}\label{sec:numerical-examples}
In this section, we show a basic comparison between some of the heuristics presented in~\S\ref{sec:optimization-problem}
and the performance bounds presented in~\S\ref{sec:performance-limits} on a small problem of designing
a Helmholtz resonator in one and two dimensions. In these cases, we can certify that the heuristics find a device whose
performance is at most $2\%$ above the global optimum, even though the original problem is likely hard.
We then show examples in the literature of much larger devices for which
no lower bound has been computed (and, indeed, cannot be computed with the current methods in a reasonable
amount of time), but has resulted in practically useful devices whose performance is much better than
that of traditional designs.

\subsection{Small examples}\label{sec:small-numerical-examples}
In this section, we will compare the performance of three heuristics, sign-flip descent (SFD), IPOPT,
and genetic algorithms (GA), and the lower bounds presented in~\S\ref{sec:performance-limits}. We will
first compare their respective performance on a small, one-dimensional design, where both the power
lower bounds and the dual lower bounds can be computed in a reasonable amount of time, and then compare
their performance on a larger two-dimensional design. The times reported in this section are from a dual-core 2015
MacBook Pro laptop running at 2.9GHz with 8GB of RAM. At the time of writing,
this machine is 5 years old and is roughly representative of the computational
power available in more recent standard and lower-end consumer laptops.

\paragraph{Problem formulation.} In both scenarios, our objective function $f$ will be to best
match a desired field $\hat z$,
\[
f(z) = \|z - \hat z\|_2^2,
\]
where the desired field is a cosine wave with a Gaussian envelope on the left half of the domain,
and is equal to zero on the right half. The excitation is a single delta function in the center of
the domain. In this problem, the designer is then allowed to choose the speed of the wave at each
point in space (via the design parameters $\theta$), such that the objective function is
minimized. (For more details and code, see the appendix~\ref{app:numerical-examples}.)

\begin{figure}
	\centering
	\includegraphics[width=.9\textwidth]{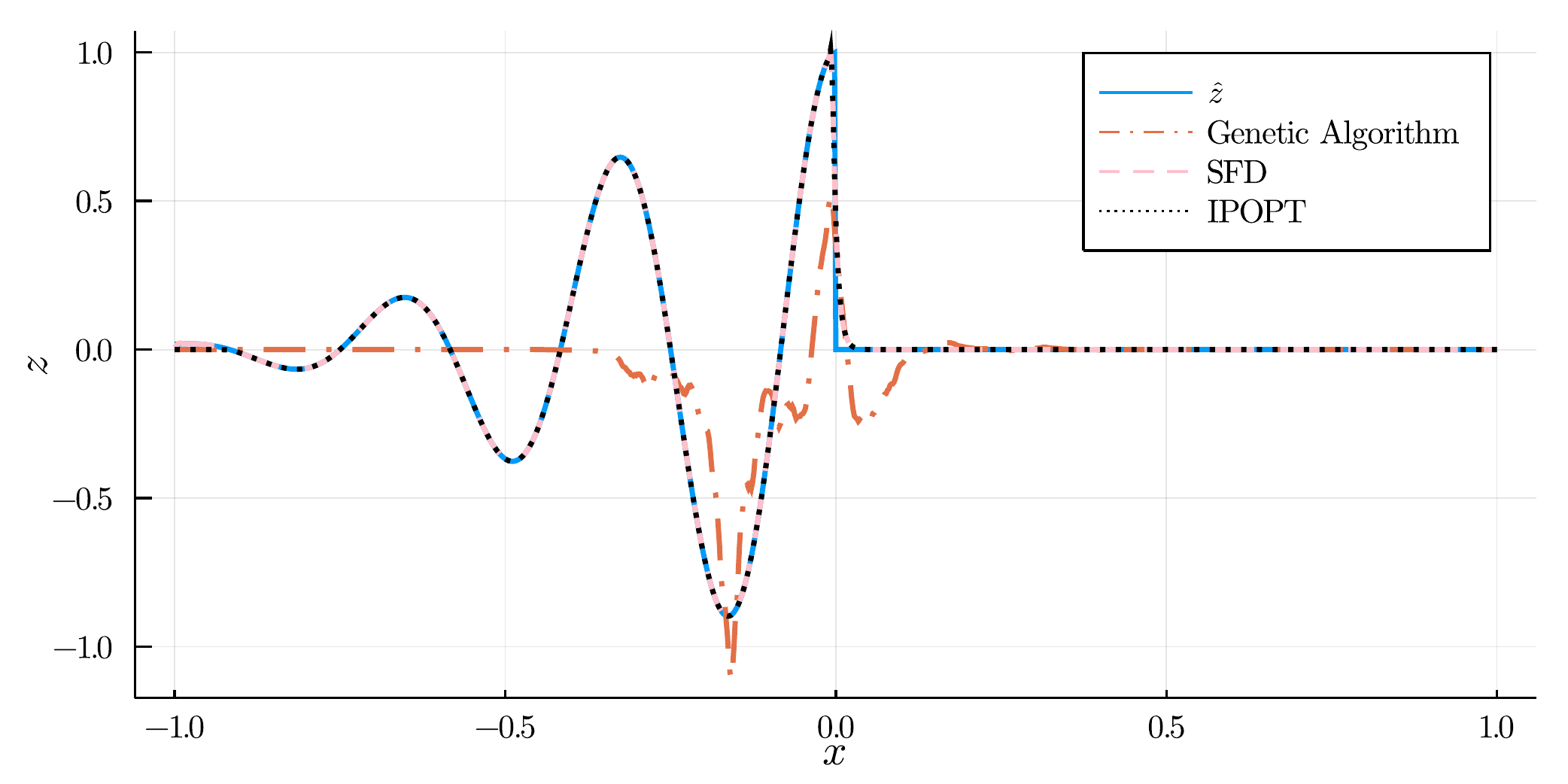}
	\caption{Approximate designs and desired field for 1D problem.}
	\label{fig:1d-comparison}
\end{figure}

\paragraph{1D problem.} We compare the objective performance and
computational performance of the heuristics and bounds in table~\ref{tbl:small-design}.
We also plot the
corresponding fields and the desired field in figure~\ref{fig:1d-comparison}.
While GA gives rather poor solutions (which only somewhat match the largest
features of the desired field), IPOPT and SFD give approximately-optimal fields
that are essentially indistinguishable from the desired one.

\begin{table}
\centering
\begin{tabular}{| c || c | c |}
\hline
 Algorithm & Objective value & Time (s) \\ 
\hline
\hline
 Genetic algorithm & 2.54 & 6.40 \\  
\hline
 IPOPT & .652 & 1.70 \\
\hline
 Sign-flip descent & .642 & .592 \\
\hline
\hline
 Power dual bound & .639 & 125\\
\hline
 Diagonal dual bound & .634 & 1.39 \\
 \hline
\end{tabular}
\caption{Performance results for small design.}
\label{tbl:small-design}
\end{table}
We find, at least in this small scenario, that IPOPT and SFD have objective
values that are extremely close to that of the power bounds and diagonal dual
bounds, which means that the designs found in this scenario can, for all
intents and purposes, be considered globally optimal. We also note that GA,
while simple to implement, does not find a good solution even with some
amount of tuning, while also having the worst performance of the available
heuristics in terms of total time taken. Additionally, while the power bound
was slightly tighter than the diagonal dual lower bound, it took nearly 90
times longer to converge for this small problem. This is due to the fact that
SDPs solution times scale approximately cubically with the problem
dimension (\ie, are $O(n^3)$), which quickly becomes an issue for larger
problems.

In fact, we note that it was difficult to find a desired field
$\hat z$ where the bounds and the performance of designs found by SFD differed
significantly. We encourage readers to search for some cases where this is true
by trying a few different desired fields in the code available for this paper.
For more details, see appendix~\ref{app:1d-problem}.

\paragraph{2D problem.} In the 2D problem, we again test the performance of GA,
IPOPT, and SFD, and the diagonal design lower bound, though we do not compute
the power bounds. We note that na\"ively attempting to compute the power bounds
in~\eqref{eq:dual-problem-family-bounds} by framing the problem as an SDP
(using an SDP solver that does not support chordal sparsity) results in a dense
matrix variable of size $(251^2)^2/2 = 251^4/2$, which requires approximately
16GB of memory to store, nearly double the available memory (8GB), not counting
the additional memory required to perform operations on this matrix.

\begin{figure}
	\centering
	\includegraphics[width=\textwidth]{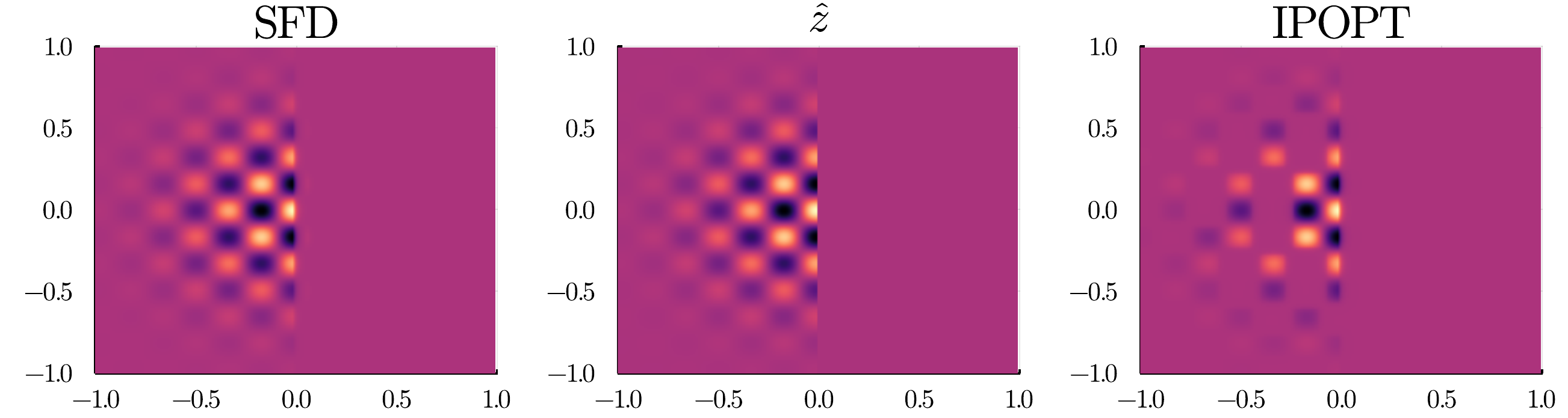}
	\caption{Approximate designs and desired field for 2D problem.}
	\label{fig:2d-comparison}
\end{figure}

The results of this comparison are available in table~\ref{tbl:large-design}
and the fields of the approximately optimized designs are shown in
figure~\ref{fig:2d-comparison}. We terminated any algorithm whose runtime was
longer than 30 minutes on the current computer. We note that, again, SFD has
surprisingly good performance, and, when combined with the lower bound, yields
a design that is guaranteed to be no more than $(11.9/11.7 - 1) \approx 1.7\%$
suboptimal, relative to the globally optimal value (in fact, as shown in
figure~\ref{fig:2d-comparison}, the resulting field from SFD and the desired
field, $\hat z$, are difficult to visually distinguish). Additionally, solving
the diagonal dual bound was faster than getting an approximate design from any
of the heuristics, though we note that this is likely not the case with much
larger designs or better-optimized heuristics. For more details, see appendix~\ref{app:2d-problem}.
\begin{table}
\centering
\begin{tabular}{| c || c | c |}
\hline
 Algorithm & Objective value & Time (s) \\ 
\hline
\hline
 Genetic algorithm & N/A & $> 1800$ \\  
\hline
 IPOPT & 190.71 & 1360 \\
\hline
 Sign-flip descent & 11.9 & 364 \\
\hline
\hline
 Diagonal dual bound & 11.7 & 48.5 \\
 \hline
\end{tabular}
\caption{Performance results for larger design.}
\label{tbl:large-design}
\end{table}

\paragraph{Discussion.} The examples in this section show that modern heuristics
have very good practical performance when compared to the available lower bounds.
In fact, we suspect that this is true more generally: modern heuristics with reasonable
initializations likely return designs that are very good, if not globally optimal, even
when there exist no bounds that can certify this, or when the available bounds
are very weak. Additionally, while we have made some basic performance optimizations to the
available code, we opted for clarity as opposed to pure performance in the
implementation of the algorithms presented, so the numbers above should only be
interpreted as general guidelines.
The code to generate the plots and tables is available at \texttt{https://github.com/cvxgrp/pd-heuristics-and-bounds}.

\subsection{Practical examples}
In this section, we show practical examples in the literature where the devices found by heuristics
have been fabricated and experimentally verified. While the current bounds
cannot be used to certify that the performance of these designs is close to
globally optimal in a reasonable amount of time, the resulting designs
have much better performance than that of traditional designs.
For a comprehensive overview of the history of inverse design and its
applications in practice, including older literature, we refer the reader
to~\cite{moleskyInverseDesignNanophotonics2018}.

%While the formulation of problem~\eqref{eq:main} may seem simple, it includes many practical
%physical design problems. In some cases, approximate solutions have been fabricated
%and experimentally verified. Here, we discuss instances in the literature relating
%specifically to photonics and show some numerical examples of the bounds
%applied to small problems.

\paragraph{Splitters.} Perhaps the most striking applications of
physical design is in the design of
compact \emph{splitters}---devices which, given some input in a specific
scenario (for example, given an input at a specific frequency) must direct as
much of the input as possible into a desired output location. Different
scenarios would direct the input to different locations; \ie, the input is
`split' to different outputs depending on the scenario. Some examples of such
devices and their performance can be found in,
\eg,~\cite{piggottInverseDesignDemonstration2015,
shenIntegratednanophotonicsPolarizationBeamsplitter2015,
makBinaryParticleSwarm2016}.

\begin{figure}
	\centering
	\includegraphics[width=.4\textwidth]{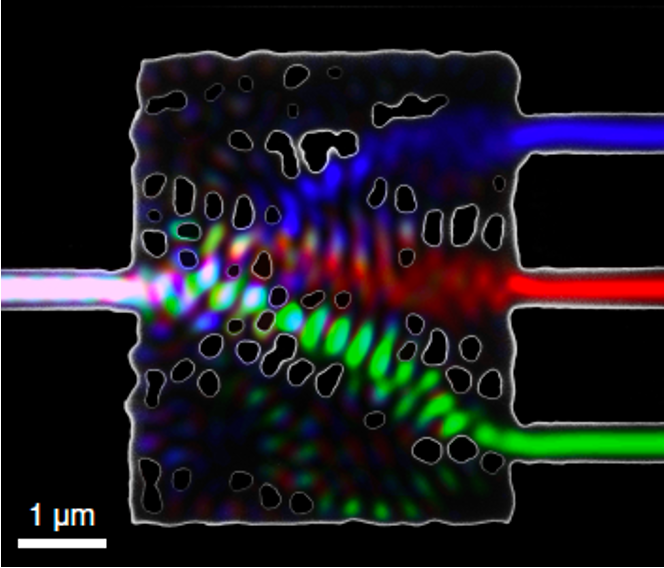}
	\caption{Scanning-electron microscope image of a three-way wavelength splitter with simulated fields overlaid. Figure and design from~\cite{suInverseDesignDemonstration2018}.}
	\label{fig:wdm}
\end{figure}

Figure~\ref{fig:wdm} shows an example
of the splitter designed and fabricated in~\cite{suInverseDesignDemonstration2018}. In this
figure, three different inputs are represented by the three different colors (blue, red,
and green), which are initially `mixed' in the input on the left-hand-side of the domain.
The device then separates the wavelengths into each of the three output channels.

\begin{figure}
	\centering
	\includegraphics[width=.4\textwidth]{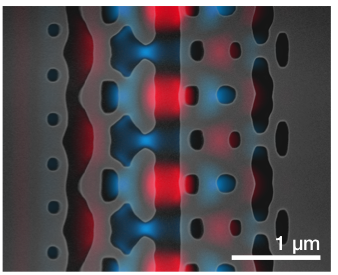}
	\caption{Scanning-electron microscope image of an inverse-designed laser-driven particle accelerator with simulated fields overlaid. Figure and design from~\cite{sapraOnchipIntegratedLaserdriven2020}.}
	\label{fig:accelerator}
\end{figure}

\paragraph{Particle accelerators.} There are other types of devices which have been designed and manufactured
in practice. A recent demonstration of a laser-driven particle accelerator
small enough to be placed on a chip~\cite{sapraOnchipIntegratedLaserdriven2020}
required several components to be efficient enough that the input power did
not destroy the material out of which the components were made. The necessary
components included the devices which coupled the laser to the structure, along
with the actual accelerator. The resulting
designs are unintuitive enough that it is not clear a human could manually
design a device of the same (or smaller) size that had at least the same efficiency.
The accelerator from the final device of~\cite{sapraOnchipIntegratedLaserdriven2020} is shown in figure~\ref{fig:accelerator}.

\paragraph{Lenses.}
There have been a few other types of devices that have been designed by these methods
and created in practice. Some examples include flat lenses with a large depth of field that have the same focusing
efficiency as traditional lenses~\cite{bayatiInverseDesignedMetalenses2020}, and
flat lenses with an ultra-wide field of view~\cite{banerjiInverseDesignedFlat2020}.
There has been some additional work in creating deformable lenses whose focal length
can be controlled by stretching or contracting the material, with performances
exceeding those of traditional lenses~\cite{callewaertInversedesignedStretchableMetalens2018}.

\paragraph{Fabrication constraints.} We note that, while there have been many numerical demonstrations of inverse designed
photonic devices, the actual fabrication of such devices has been relatively difficult
until recently, when methods that could include fabrication
constraints~\cite{piggottFabricationconstrainedNanophotonicInverse2017,
vercruysseAnalyticalLevelSet2019} and
could accommodate a large enough number of field variables became available
and were fast enough to be used in practice.
Additionally, fabrication constraints can also be applied to mass-produced photonic
circuits in a foundry~\cite{piggottFoundryfabricatedInverseDesigned2019},
which makes it possible to fabricate inverse-designed photonics at scale.

\section{Future directions}
While physical design and, more specifically, inverse design, has led to some dramatic
improvements in the performance of photonic devices, there are still many questions left
unanswered and possible future avenues for research.

\paragraph{Standard benchmarks.} There is a need for
standardized benchmarks for both heuristic algorithms and bounds. Similar to
the standard performance benchmarks for machine
learning (such as MNIST~\cite{lecunGradientbasedLearningApplied1998},
ImageNet~\cite{dengImageNetLargescaleHierarchical2009},
CIFAR-10~\cite{krizhevskyLearningMultipleLayers2009}, \etc) and for
optimization (such as the Maros--M\'esz\'aros test
set~\cite{marosRepositoryConvexQuadratic1999}, MIPLIB
2017~\cite{gleixnerMIPLIB2017DataDriven}, \etc), which are used to
compare the performance of different proposed algorithms, it is feasible
to have a standard library of design specifications (objective functions and
constraints) which either have known global solutions or a best known solution
and best known lower bound that is updated as better ones are found. This would
allow for researchers to have a concrete library which can be used to compare
proposed algorithms against existing ones, in terms of both objective value and
computational performance. Like in machine learning or optimization, we suspect
that such benchmarks would clarify the performance tradeoffs of different
algorithms and bounds. These benchmarks would help identify what
problems are `hard' for current heuristics and possibly lead to better
approaches.

\paragraph{Improved heuristics.} While some of the methods presented here can
scale to very large designs with millions to tens of millions of field and tens
of thousands of design variables, many of the current methods do not make use
of the specific structure of the physical design problem and take a long
time (days or weeks) even with large computers or clusters. For example, one
possible avenue is the use of a primal-dual algorithm, which can lead to
drastically improved performance for some problems (see,
\eg,~\cite[\S3.5]{oconnorPrimalDualDecompositionOperator2014}). Such
methods could also lead to algorithms whose natural byproduct of computing
an approximately optimal design is a bound, like
in~\S\ref{sec:performance-limits}, guaranteeing that the resulting design is
close to the global optimum without requiring additional computational time.
There has been additional work with combining machine learning approaches
to speeding up both simulations~\cite{trivediDatadrivenAccelerationPhotonic2019}
and optimization~\cite{liuTrainingDeepNeural2018,
jiangGlobalOptimizationDielectric2019, soDeepLearningEnabled2020, chenGenerativeDeepNeural2020},
which trade off training time for runtime performance.

\paragraph{Improved initializations.} All of the heuristics used in practice
often require good initializations in order to have reasonable performance.
In practice, initializations that are guided by intuition appear to work
well for many scenarios, but sometimes fail to reach what are known
to be better designs. (See, \eg,~\cite[\S5]{suNanophotonicInverseDesign2020}.)
As previously discussed, the bounds presented yield good initial designs
as a by-product of the optimization procedure; but this need not be true
in general, and we suspect that there might exist better methods which,
while expensive to run exactly, might yield good initial designs. For example,
while sign-flip descent might be potentially very expensive to run for large
designs (as it requires solving a general constrained convex problem at each iteration)
one could perform a small number of iterations to get a reasonably
good initial field and then feed this resulting field into an optimization
algorithm with faster convergence. (This latter procedure is sometimes called
`solution polishing' in combinatorial optimization.) Another avenue of
similar work is the idea of `objective-first' optimization~\cite{luObjectivefirstDesignHighefficiency2012}
where the physics equation is relaxed to a penalty with a small weight and
the resulting nonconvex problem is solved to get an infeasible design with
good objective performance and (hopefully) small physics residual. This initial, infeasible
design is then passed to any of the heuristics above to attempt to find a
feasible design and field with similar performance.

\paragraph{Improved bounds.} The small examples we have discussed here are simple
cases where the bounds and the heuristics are very close; but this need not be
the case. For example, the diagonal dual bounds given here can give weak bounds
when the objective function depends only on a few entries of the field variable
$z$, even with the additional extensions presented. While we, in general,
suspect the power bound yields tighter results than the diagonal dual lower
bound in these cases, current off-the-shelf solvers are too slow to solve the
resulting problem other than for a small number of field variables, $n \lesssim 10^3$.
One future research avenue is to create faster solvers for
problem~\eqref{eq:dual-problem-family-bounds}, by exploiting the specific
structure available in these problems. A second possibility is to find some connection between
the diagonal dual and power bounds; in particular, it is not clear how one is
related to the other, if at all. We conjecture that the power
bounds~\eqref{eq:dual-problem-family-bounds} are always guaranteed to be at
least as tight as the diagonal dual ones~\eqref{eq:diagonal-dual} whenever the
function $f$ is a separable convex quadratic, and where the constraint sets
$S_k$ are singletons (equation~\eqref{eq:constraint-set} with $S_k = \{k\}$ for
$k=1, \dots, n$) but have been unable to prove this.
It is also of practical interest to create lower bounds which give approximate
scaling rules for designs, which would help a designer choose appropriate
device sizes and materials for a desired objective.

\section{Conclusion}
%While there has been a large amount of recent research in the area
%of physical design, leading to interesting experimental and theoretical results,
%physical design (or more specifically, inverse design) is not yet a
%\emph{technology}; \ie, a collection of methods and software requiring nearly
%no user input apart from the basic problem specification. While current solvers
%and software work very well for many applications, the problem
%specification is tricky and often requires careful experimentation. (For
%example, many heuristics require good initializations to find devices with
%reasonable performance.) Currently, there are no
%general benchmarks that can be used to compare the computational performance or
%objective performance of different heuristics, which makes it difficult to
%asses the general performance of algorithms, other than in specific scenarios.
%Yet, despite the drawbacks of the current approaches, we suspect that physical
%design will, in the near future, become a standard way of creating
%practical, efficient devices that far exceed the performance of their
%traditionally-designed counterparts.

There has been tremendous progress in the area of photonic inverse design,
including foundry-based fabrication and the use of design software in
a wide range of academic and industry labs. Still, there is a lot to be improved,
including, but not limited to: designing standard benchmarks and comparisons for
heuristics, overcoming computational bottlenecks in order to design larger devices
than those which are currently possible, and
the improvement of
bounds (both theoretically and in terms of computational performance) which currently
either do not apply, or are very difficult to compute, for many of the devices used in
practice. Yet, despite the drawbacks of the current approaches, inverse design
has yielded incredible practical benefits, and we expect that such methods will,
in the near future, become a standard approach to creating
practical, efficient devices that far exceed the performance of their
traditionally-designed counterparts.

\section*{Acknowledgements}
The authors would like to thank Rahul Trivedi for useful discussions regarding
the current literature on bounds and Carlos G.\ Gonzalez for comments and corrections.
The authors would also like to acknowledge
financial support from ARPA-E with Agreement No.\ DE-AR0001212. Guillermo Angeris is supported by the
National Science Foundation Graduate Research Fellowship under Grant No.\
DGE-1656518.

\bibliographystyle{ieeetr}
\bibliography{citations.bib}

\clearpage

\appendix

\section{Bound equivalence}
\label{app:bound-equivalence}
We show that the bounds presented in~\cite{moleskyHierarchicalMeanFieldOperator2020, kuangComputationalBoundsLightmatter2020} are equivalent to the ones given in section~\ref{sec:local-power-conservation}.

\paragraph{Derivation.} First, we derive the bounds presented using this notation. We define the \emph{displacement field}
$y_i = \gamma_iz_i$, for $i=1,\dots, n$, where $\gamma_i = (1+\theta_i)/2$. (Note that $\theta_i \in [-1, 1]$ whenever $\gamma_i \in [0, 1]$.) The diagonal physics equation can then be written
in terms of the field $z$ and the displacement field $y$:
\[
(2A_0 - I)z + y = 2b_0, \qquad y_i = \gamma_i z_i, \qquad i=1, \dots, n,
\]
where $\gamma \in [0, 1]^n$. We will define $G = (2A_0 - I)^{-1}$ (this can be recognized, roughly speaking, as the Green's function) and $b' = 2Gb_0$ (which is sometimes called the `incident field') for notational convenience. We can then rewrite the diagonal physics equation using $G$ and $b'$:
\begin{equation}\label{eq:greens-formalism}
z_i + g_i^Ty = b_i', \qquad y_i = \gamma_i z_i, \qquad i=1, \dots, n,
\end{equation}
where $g_i^T$ denotes the $i$th row of $G$. Multiplying on the left by $y_i$, we find that $y$ and $z$ must satisfy
\[
y_i(z_i + g_i^Ty) = y_i b_i', \quad i=1, \dots, n.
\]
Note that $y_i^2 = \gamma_i^2 z_i^2 \le \gamma_i z_i^2 = y_iz_i$ because $\gamma_i^2 \le \gamma_i$, since $\gamma_i \in [0, 1]$. Using this result, we find the following quadratic inequalities depending only on the displacement field $y$:
\[
y_i(y_i + g_i^Ty) \le y_i b_i', \quad i=1, \dots, n.
\]
Scaling each of the $i$ inequalities by any nonnegative value $\lambda_i \ge 0$ and summing them together implies that $y$ must satisfy:
\begin{equation}\label{eq:displacement-bound}
	y^TDy + y^TDGy \le y^TDb',
\end{equation}
where $D = \diag(\lambda)$ is any matrix with nonnegative diagonal entries. These are the bounds presented 
in~\cite{moleskyHierarchicalMeanFieldOperator2020, kuangComputationalBoundsLightmatter2020} in the case where
$\Theta$ is a box rather than a Boolean constraint.

\paragraph{Tightness.} As expected, these bounds are also tight in the sense that, if $y$ satisfies~\eqref{eq:displacement-bound}
for all nonnegative diagonal matrices $D$, then there exists a feasible design $\theta$ and a field $z$ such that $z$ and $\theta$ 
satisfy the diagonal physical equation, and the displacement field $y$ satisfies $y_i = (1+\theta_i)z_i/2$ for $-1\le\theta_i\le 1$, or, 
equivalently, that $y_i = \gamma_iz_i$, where $0 \le \gamma_i \le 1$ for $i=1, \dots, n$.

To show this, we will consider (as before) only
the diagonal matrices $D = e_ie_i^T$. The bound then implies that
\[
y_i(y_i + g_i^Ty) \le y_i b_i', \quad i=1, \dots, n.
\]
We can then choose $z_i = b_i' -g_i^Ty$ and
\[
\gamma_i = \begin{cases}
y_i/(b_i' - g_i^Ty) & b_i' - g_i^Ty \ne 0\\
0 & \text{otherwise},
\end{cases}
\]
for $i=1, \dots, n$. Note that $z_i$ obviously satisfies $z_i + g_i^Ty = b_i'$ and, whenever $b_i' - g_i^Ty \ne 0$, we have $y_i = \gamma_iz_i$, by construction. On the other hand, if $b_i' - g_i^Ty = 0$, then
\[
y_i^2 = y_i(y_i + g_i^Ty - b_i') \le 0,
\]
so $y_i = 0 = \gamma_iz_i$, since $\gamma_i = 0$ whenever $b_i' - g_i^Ty = 0$, by definition. In other words, given a displacement field 
$y$ satisfying~\eqref{eq:displacement-bound} for all nonnegative diagonal matrices, we have constructed a field $z$ and a design $\theta 
= 2\gamma - \ones$ such that $y$, $z$, and $-\ones \le \theta \le \ones$ all satisfy the physics equation~\eqref{eq:greens-formalism}, 
or, equivalently, $z$ and $\theta$ satisfy the diagonal physics equation.

%For each $i=1, \dots, n$, either the inequality holds strictly or it holds at equality. If the $i$th inequality holds strictly, note that
%\[
%y_i(y_i + g_i^Ty - b_i') < 0
%\]
%implies
%\[
%y_i > 0, \quad y_i + g_i^Ty - b_i' < 0, \qquad \text{or} \qquad y_i < 0, \quad y_i + g_i^Ty - b_i' > 0.
%\]
%Because of this, we define
%\[
%\alpha_i = -\frac{y_i + g_i^Ty - b_i'}{y_i} > 0,
%\]
%such that
%\[
%(1+\alpha_i)y_i + g_i^Ty - b_i' = 0,
%\]
%for each $i=1, \dots, n$. If we set $\gamma_i = 1/(1+\alpha_i)$ then $y_i = z_i/(1 + \alpha_i)$ and $z_i + g_i^Ty = b_i'$ (so $y_i$ and $z_i$ satisfies the $i$th entry of the physics equations). Additionally, $0 < \gamma_i < 1$ since $\alpha_i > 0$.
%
%On the other hand, if the inequality holds at equality, we then have that either
%$y_i = 0$ or $y_i + g_i^Ty - b_i' = 0$. If $y_i=0$, we can set $\gamma_i = 0$ and, we can choose any $z_i$, so we set it to satisfy the $i$th entry of the physics equation $z_i = -(g_i^Ty - b_i')$. On the other hand,
%if $y_i + g_i^Ty - b_i' = 0$ then we set $\gamma_i = 1$ such that $z_i = y_i$. In either case, $z_i + g_i^Ty - b_i' = 0$,
%and so $z$ satisfies the physics equation, while $0 \le \gamma \le \ones$, as required.

\paragraph{Equivalence.} The equivalence between the bounds follows immediately from the fact that~\eqref{eq:displacement-bound}
holds for all nonnegative diagonal matrices $D$ if, and only if, the physics equation~\eqref{eq:greens-formalism} holds for this choice 
of $y$, $z$, and $\theta$, which, in turn, holds if, and only if, the original family of power bounds over $z$ and $\theta$ holds for 
all diagonal matrices.

\section{Numerical examples}\label{app:numerical-examples}
In this section, we focus on two basic numerical experiments for comparing both
heuristics and lower bounds.
Additionally, all Julia~\cite{bezansonJuliaFreshApproach2017} code for the
examples can be found at
\texttt{https://github.com/cvxgrp/pd-heuristics-and-bounds}. The optimization
problems used in computing the bounds were specified using the
JuMP~\cite{dunningJuMPModelingLanguage2017} domain-specific language in Julia,
and solved using Mosek~\cite{mosek}.

\subsection{Small design}\label{app:1d-problem}
In this scenario, we are given a device whose field must satisfy the 
discretized Helmholtz equation in one dimension. At every point
in the domain, the designer is allowed to choose the speed of the wave
in the material within some specified range, such that the resulting
field best matches some desired field.

\paragraph{Helmholtz's equation.} In one dimension, we can write
Helmholtz's equation in the interval $[-1, 1]$ as:
\begin{equation}\label{eq:helmholtz-1d}
\frac{\partial^2 u(x)}{\partial x^2} + \frac{\omega^2}{c(x)^2}u(x) = v(x), \quad -1 \le x \le 1,
\end{equation}
where $u: [-1, 1] \to \reals$ is the amplitude of the wave, while $\omega \in
\reals$ is the angular frequency and $c:[-1, 1] \to \reals_{++}$ is the speed
of the wave in the material, and $v: [-1, 1] \to \reals$ is the excitation. We
will assume Dirichlet boundary conditions for simplicity; \ie, that $u(-1) =
u(1) = 0$.

\paragraph{Discretization.} We can write a simple discretization of the above
equation by subdividing the interval $[-1, 1]$ into $n$ equidistant points
$\{x_i\}$ for $i=1, \dots, n$. We then approximate the second derivative using
finite-differences such that
\[
\frac{\partial^2u(x_i)}{\partial x^2} \approx \frac{z_{i-1} - 2z_i + z_{i+1}}{h^2}, \quad i=1, \dots, n,
\]
where $h$ is the separation $x_i - x_{i-1}$ for any $i=2, \dots, n$ and is
equal to $h=2/(n-1)$, while $z_i = u(x_i)$ for $i=1, \dots, n$. Additionally,
we have defined $z_0 = z_{n+1} = 0$ for convenience.

 If we then define $\omega^2/c(x_i)^2 = \theta_i$, then we can approximate
 equation~\eqref{eq:helmholtz-1d} as a diagonal physics equation of the
 form~\eqref{eq:diagonal-design}, where
\begin{equation}\label{eq:second-deriv-approximation}
A_0 = \frac{1}{h^2}\begin{bmatrix}
	-2 & 1 & 0 & \dots & 0\\
	1 & -2 & 1 & \dots & 0\\
	0 & 1 & -2 & \dots & 0\\
	\vdots & \vdots & \vdots & \ddots & \vdots\\
	0 & 0 & 0 & \dots & -2
\end{bmatrix},
\end{equation}
and $(b_0)_i = v(x_i)$ for $i=1, \dots, n$.

\paragraph{Problem data.} In this problem, we will seek to minimize the squared
Euclidean distance between the field $z$ and some desired field $\hat z$; \ie,
the objective is
\[
f(z) = \|z - \hat z\|_2^2 = \sum_{i=1}^n (z_i - \hat z_i)^2,
\]
and we choose $\hat z$ to be a cosine wave with a Gaussian envelope of width
$\sigma/\sqrt{2}$, whenever $x_i < 0$ and $0$ when $x_i \ge 0$:
\begin{equation}\label{eq:desired-z-1d}
\hat z_i = \begin{cases}
\cos(\omega x_i)\exp(-x^2_i/\sigma^2) & x_i < 0\\
0 & x_i \ge 0.
\end{cases}
\end{equation}
Note that this function is discontinuous at 0. For the small numerical
experiment, we will choose $\sigma = 1/2$, $\omega = 6\pi$, and $\Theta = [1,
1.5]^n$. We will also set $b_{(n+1)/2} = 2$ and zero otherwise as our
excitation, and set $n$, the number of gridpoints, to be $n=1001$. This means
that there are $1001$ design parameters and field variables, which, in
practice, is a very small number when compared to current applications.

\subsection{Larger design}\label{app:2d-problem}
In this example, we will show an example of a larger physical design problem
with $n = d = 251^2 = 63001$, that is similar in spirit to the smaller design,
but is large enough that sparsity has to be exploited in order to have
reasonable run time performance.

\paragraph{2D Helmholtz equation.} For this problem, we discretize the 2D
Helmholtz's equation:
\[
\frac{\partial^2 u(x, y)}{\partial x^2} + \frac{\partial^2 u(x, y)}{\partial y^2} + \frac{\omega^2}{c(x, y)^2}u(x, y) = v(x, y),
\]
over the domain $(x, y) \in [-1, 1]^2$. Here, as before $u: [-1, 1]^2 \to
\reals$ is the amplitude of the wave, while $\omega \in \reals$ is the angular
frequency, $c: [-1, 1]^2 \to \reals_{++}$ is the speed of the wave in the
material, and $v: [-1, 1]^2 \to \reals$ is the excitation. We assume Dirichlet
boundary conditions; \ie, that $u(x, y) = 0$ at the boundary of the domain.
\paragraph{Discretization.} Assuming that we have $n=l^2$ equally spaced
gridpoints $(x_i, y_i) \in [-1, 1]^2$ (\ie, there are $l$ points along a given
axis), we will let $z_i$ be the approximate value of $u(x_i, y_i)$. Writing the
second-order difference matrix in~\eqref{eq:second-deriv-approximation} as
$\Delta \in \reals^{l\times l}$, we can approximate the sum of the partial
derivatives as
\[
\frac{\partial^2 u(x_i, y_i)}{\partial x^2} + \frac{\partial^2 u(x_i, y_i)}{\partial y^2} \approx ((I_l \otimes \Delta + \Delta \otimes I_l)z)_i, \quad i=1, \dots, n,
\]
where $I_l$ is the $l\times l$ identity matrix, and $\otimes$ is the Kronecker
product. The resulting problem is then a diagonal physical design problem with
\[
A_0 = I_l \otimes \Delta + \Delta \otimes I_l,
\]
and $(b_0)_i = v(x_i, y_i)$ for $i=1, \dots, n$. Note that the size of $A_0$ is
$n \times n = l^2 \times l^2$; so the total number of entries of the matrix
grows in the fourth power of the side length. On the other hand, the resulting
matrix is very sparse because less than $5n$ of the entries are not zero.
In other words, when $l=100$, the percentage of nonzero entries is less than
$.05\%$.

\paragraph{Problem data.} As before, the objective is to minimize the squared
difference of the resulting field $z$ and some desired field $\hat z$:
\[
f(z) = \|z - \hat z\|_2^2.
\]
Here $\hat z$ is the two-dimensional analogue of~\eqref{eq:desired-z-1d},
\[
\hat z_i = \begin{cases}
	\cos(\omega x_i) \cos(\omega y_i) \exp(-(x_i^2 + y_i^2)/\sigma^2) & x_i < 0\\
	0 & x_i \ge 0,
\end{cases}
\]
for $i=1, \dots, n$. We will again choose $\Theta = [1, 1.5]^n$, $\omega =
6\pi$ and $\sigma = 1/2$, and $b_{(n^2+1)/2} = 1$, along with $l=251$ (and
$n=l^2 = 65001$), which results in a problem with $65001$ design parameters and
field variables. We encourage the reader to try different choices parameters in
the provided code and explore the resulting heuristic performance, bounds, and
time taken by the algorithms.

\end{document}